\documentclass[longauth]{aa}  
\pdfoutput=1
\usepackage{graphicx}
\usepackage{txfonts}
\usepackage{natbib}
\usepackage[colorlinks=true, citecolor=blue]{hyperref}
\bibpunct{(}{)}{;}{a}{}{,}

\newcommand{\orcid}[1]{\href{https://orcid.org/#1}{\includegraphics[width=10pt]{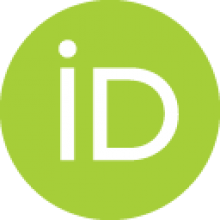}}}

\begin{document} 

\title{APOGEE discovery of a chemically atypical star disrupted from NGC~6723 and captured by the Milky Way bulge}
		
\author{
Jos\'e G. Fern\'andez-Trincado\inst{1,2}\thanks{To whom correspondence should be addressed; E-mail: jose.fernandez@uda.cl and/or jfernandezt87@gmail.com}\orcid{0000-0003-3526-5052},
Timothy C. Beers\inst{3}\orcid{0000-0003-4573-6233},
Dante Minniti\inst{4,5}\orcid{0000-0002-7064-099X},
Leticia Carigi\inst{6},
Vinicius M. Placco\inst{7}\orcid{0000-0003-4479-1265},
Sang-Hyun Chun \inst{8},
Richard R. Lane\inst{1},
Doug Geisler\inst{9,10,11},
Sandro Villanova\inst{9},
Stefano O. Souza\inst{12}\orcid{0000-0001-8052-969X},
Beatriz Barbuy\inst{12}\orcid{0000-0001-9264-4417},
Angeles P\'erez-Villegas\inst{13}\orcid{0000-0002-5974-3998},
Cristina Chiappini\inst{14,15},
Anna. B. A. Queiroz\inst{14},
Baitian Tang\inst{16},
Javier Alonso-Garc\'ia\inst{17,18},
Andr\'es E. Piatti\inst{19,20}\orcid{0000-0002-8679-0589},
Tali Palma\inst{21,22},
Alan Alves-Brito\inst{23},
Christian Moni Bidin\inst{24},
Alexandre Roman-Lopes\inst{10}\orcid{0000-0002-1379-4204},
Ricardo R. Mu\~noz\inst{25,26},
Harinder P. Singh\inst{27},
Richa Kundu\inst{27,28},
Leonardo Chaves-Velasquez\inst{29,30}\orcid{0000-0002-9677-1015},
Mar\'ia Romero-Colmenares\inst{17},
Penelope Longa-Pe\~na\inst{17},
Mario Soto\inst{1}
and
Katherine Vieira\inst{1}\orcid{0000-0001-5598-8720}
}
	
	\authorrunning{Jos\'e G. Fern\'andez-Trincado et al.} 
	
\institute{
         Institut Utinam, CNRS UMR 6213, Universit\'e Bourgogne-Franche-Comt\'e, OSU THETA Franche-Comt\'e, Observatoire de Besan\c{c}on, \\ BP 1615, 25010 Besan\c{c}on Cedex, France
         \and 
         Instituto de Astronom\'ia y Ciencias Planetarias, Universidad de Atacama, Copayapu 485, Copiap\'o, Chile
         \and
         Department of Physics and JINA Center for the Evolution of the Elements, University of Notre Dame, Notre Dame, IN 46556, USA
         \and 
         Depto. de Cs. F\'isicas, Facultad de Ciencias Exactas, Universidad Andr\'es Bello, Av. Fern\'andez Concha 700, Las Condes, Santiago, Chile
         \and
         Vatican Observatory, V00120 Vatican City State, Italy
         \and
         Instituto de Astronom\'ia, Universidad Nacional Aut\'onoma de M\'exico, A.P. 70-264, 04510, Ciudad de M\'exico, Mexico
         \and
         NSF's National Optical-Infrared Astronomy Research Laboratory, Tucson, AZ 85719, USA
         \and
         Korea Astronomy and Space Science Institute, 776 Daedeokdae-ro, Yuseong-gu, Daejeon 34055, Republic of Korea
         \and
         Departamento de Astronom\'i a, Casilla 160-C, Universidad de Concepci\'on, Concepci\'on, Chile
         \and
         Departamento de Astronom\'ia, Universidad de La Serena, 1700000 La Serena, Chile
         \and
         Instituto de Investigaci\'on Multidisciplinario en Ciencia y Tecnolog\'ia, Universidad de La Serena. Benavente 980, La Serena, Chile
         \and
         Universidade de S\~ao Paulo, IAG, Rua do Mat\~ao 1226, Cidade Universit\'aria, S\~ao Paulo 05508-900, Brazil
         \and
         Instituto de Astronom\'ia, Universidad Nacional Aut\'onoma de M\'exico, Apdo. Postal 106, 22800 Ensenada, B.C., M\'exico
         \and
         Leibniz-Institut f\"ur Astrophysik Potsdam (AIP), An der Sternwarte 16, 14482 Potsdam, Germany
         \and
         Laborat\'orio Interinstitucional de e-Astronomia - LIneA, Rua Gal. Jos\'e Cristino 77, Rio de Janeiro, RJ - 20921-400, Brazil
         \and
         School of Physics and Astronomy, Sun Yat-sen University, Zhuhai 519082, China
         \and
          Centro de Astronom\'ia (CITEVA), Universidad de Antofagasta, Av. Angamos 601, Antofagasta, Chile
         \and
         Millennium Institute of Astrophysics, Santiago, Chile
         \and
         Instituto Interdisciplinario de Ciencias B\'asicas (ICB), CONICET-UNCUYO, Padre J. Contreras 1300, M5502JMA, Mendoza, Argentina
         \and
         Consejo Nacional de Investigaciones Cient\'{\i}ficas y T\'ecnicas (CONICET), Godoy Cruz 2290, C1425FQB,  Buenos Aires, Argentina
         \and
         Universidad Nacional de C\'ordoba, Observatorio Astron\'omico de C\'ordoba, Laprida 854, 5000 C\'ordoba, Argentina
         \and
         Consejo Nacional de Investigaciones Cient\'ificas y T\'ecnicas (CONICET), Godoy Cruz 2290, Ciudad Aut\'onoma de Buenos Aires, Argentina
         \and 
         Universidade Federal do Rio Grande do Sul, Instituto de F\'isica, Av. Bento Gon\c{c}alves 9500, Porto Alegre, RS, Brazil
         \and
         Instituto de Astronom\'ia, Universidad Cat\'olica del Norte, Av. Angamos 0610, Antofagasta, Chile
         \and
         Departamento de Astronom\'ia, Universidad de Chile, Camino del Observatorio 1515, Las Condes, Santiago, Chile
         \and
         Visiting astronomer, Cerro Tololo Inter-American Observatory, National Optical Astronomy Observatory, which is operated by the Association of Universities for Research in Astronomy (AURA) under a cooperative agreement with the National Science Foundation, Chile
         \and
         Department of Physics and Astrophysics, University of Delhi, Delhi-110007, India
         \and
         European Southern Observatory, Alonso de C\'{o}rdova 3107, 7630000 Vitacura, Santiago, Chile
         \and
         University of Nari\~no Observatory, Universidad de Nari\~no, Sede VIIS, Avenida Panamericana, Pasto, Nari\~no, Colombia
         \and
         Departamento de F\'isica de la Universidad de Nari\~no, Torobajo Calle 18 Carrera 50, Pasto, Nari\~no, Colombia
   }
	
	\date{Received: XX/XX/XXXX; Accepted: XX/XX/XXXX}
	\titlerunning{A chemically atypical star being tidally disrupted from NGC~6723}
	
	
	\abstract
	{
The central (`bulge') region of the Milky Way is teeming with a significant fraction of mildly metal-deficient stars with atmospheres that are strongly enriched in cyanogen ($^{12}$C$^{14}$N). Some of these objects, which are also known as nitrogen-enhanced stars, are hypothesised to be relics of the ancient assembly history of the Milky Way. Although the chemical similarity of nitrogen-enhanced stars to the unique chemical patterns observed in globular clusters has been observed, a direct connection between field stars and globular clusters has not yet been proven. In this work, we report on high-resolution, near-infrared spectroscopic observations of the bulge globular cluster NGC~6723, and the serendipitous discovery of a star, 2M18594405$-$3651518, located outside the cluster (near the tidal radius) but moving on a similar orbit, providing the first clear piece of evidence of a star that was very likely once a cluster member and has recently been ejected.  Its nitrogen abundance ratio ([N/Fe]$\gtrsim + 0.94$) is well above the typical Galactic field-star levels, and it exhibits noticeable enrichment in the heavy $s$-process elements (Ce, Nd, and Yb), along with  moderate carbon enrichment; all characteristics are known examples in globular clusters. This result suggests that some of the nitrogen-enhanced stars in the bulge likely originated from the tidal disruption of globular clusters.
	}
	 
	\maketitle
	
\section{Introduction} 
\label{section1}

\begin{figure*}
	\begin{center}
		\includegraphics[width=180mm]{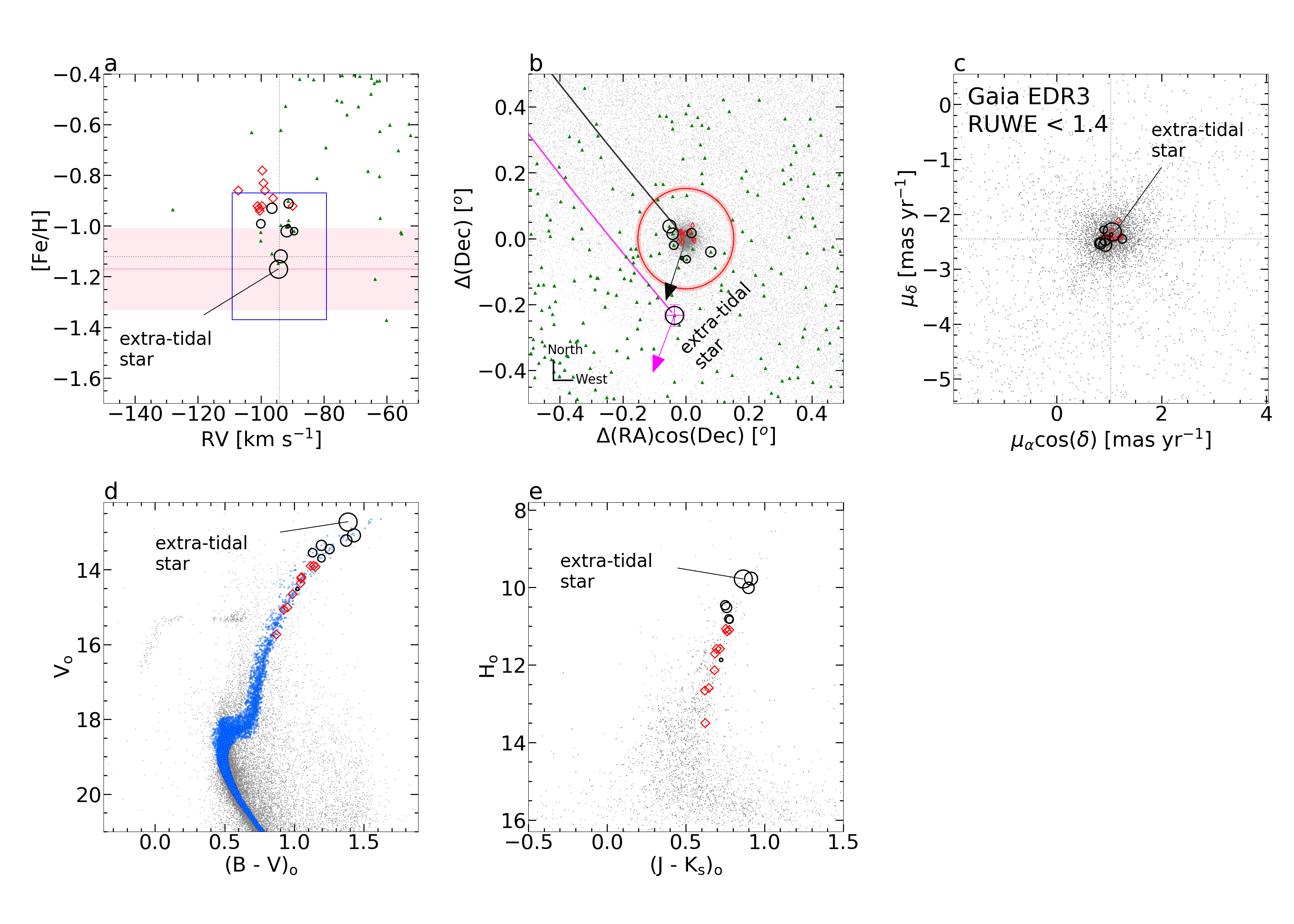}
		\caption{Properties of the extra-tidal star compared with likely members of NGC~6723. Panel (a): The metallicity -- radial velocity plane for stars in the APOGEE-2 catalogue in the region of NGC~6723 (green triangles). The black open circles are potential members of NGC~6723; sizes reflect their ${\rm V_o}$ magnitudes, with decreasing size for fainter magnitudes. The pink line and shaded region indicate the $\langle$[Fe/H]$\rangle \pm \sigma_{\rm [Fe/H]}$ of the extra-tidal star. The black dotted lines indicate the nominal [Fe/H] and radial velocity of NGC~6723 from \citet[][2010 edition]{Harris1996}. The red unfilled diamonds refer to probable members analysed in the literature \citep{Crestani2019}. The blue rectangle indicates the region adopted to choose potential cluster members, based on radial velocity and metallicity (see Section \ref{section2}). Panel (b): Positions for stars in the APOGEE-2 and \textit{Gaia} EDR3 (grey dots) catalogues in the region of NGC~6723. The magenta cross symbol indicates the \textit{Gaia} astrometric uncertainty. The big red and shadow circumference show the cluster tidal radius ($r_{t}= 9.14\pm0.49$ \arcmin) of NGC~6723 determined in this work from HST $+$ \textit{Gaia} EDR3 data set (see Section \ref{section7}). The orbital paths of the cluster (black line) and the extra-tidal star (magenta line) are shown, along with the proper motion vectors (black arrow) of the cluster from \citet[][]{Baumgardt2019}, and the extra-tidal star (magenta arrow); their lengths (scaled up for visibility) and directions are essentially identical. Panel (c): Proper-motion plane for the selected candidates. Panels (d) and (e): V$_{\rm o}$ versus (B-V)$_{\rm o}$ and 2MASS H$_{\rm o}$ versus (J-K$_{\rm s}$)$_{\rm o}$ Colour-Magnitude Diagram (CMD) centred on NGC~6723, showing all the stars within the cluster tidal radius, and the probable cluster members (cyan symbols) within a 2$\sigma$ deviation from the best isochrone fitting \citep{Souza2020, Oliveira2020}. The largest open circle in all panels indicate the extra-tidal star.}
		\label{Figure1}
	\end{center}
\end{figure*}	

The great majority of stars in the halo of the Milky Way (MW) have elemental-abundance ratios, spanning a wide range of metallicities ($-7.5 < {\rm [Fe/H]} < 0.0$) that are similar to one another and overall track the general level of metallicity. However, there are a number of important exceptions, including carbon-enhanced ([C/Fe] $\gtrsim+0.7$) metal-poor (CEMP) stars, in particular at very low metallicities ([Fe/H] $< -2.0$)--\citep[][]{Frebel2006, Lee2013, Placco2014, Placco2018, Yoon2019} and the class of CH stars (with $-2.0< $[Fe/H]$<-0.2$)--\citep{Karinkuzhi2014}, as well as the carbon-deficient stars (with [C/Fe]$\lesssim+0.5$, $-2 < {\rm [Fe/H]} < +0.1$) with light-element abundances (e.g. N, Al, and Si) whose chemical compositions mimic the abundance patterns observed in some of the stars in Galactic (and extragalactic) globular cluster (GC) populations \citep[e.g.][]{Fernandez-Trincado2017, Schiavon2017, Bekki2019, Hanke2020, Meszaros2020, Fernandez-Trincado2020d}. 

Multiple potentially pieces of evidence support the origin of CEMP stars in low-mass galaxies that have been accreted by the MW \citep{Yoon2019}, either due to mass-transfer binaries \citep[CEMP-$s$ stars,][]{Beers2005} or from natal gas polluted by high-mass stars in the early Galaxy \citep[CEMP-no stars,][]{Beers2005}. However, the mildly metal-poor giants with stellar atmospheres that are strongly enriched in $^{12}$C$^{14}$N (in cases with available data, with distinctive  Al and Si abundances as well), are referred to as nitrogen-enhanced (N-rich) stars \citep{Fernandez-Trincado2016a, Fernandez-Trincado2017, Schiavon2017, Fernandez-Trincado2019c} or NRS \citep{Bekki2019}, and defined in \citet{Johnson2007} as nitrogen-enriched metal-poor (NEMP) stars, which are CEMP stars with [C/N]$<-0.5$ and [N/Fe]$>+0.5$. They have resisted a unifying explanation of the nucleosynthetic processes responsible for their abundance anomalies. These stars share chemical-abundance patterns that are hypothesised to be associated with the so-called second-generation stars in GCs \citep[see e.g.][]{Schiavon2017b, Fernandez-Trincado2019d, Fernandez-Trincado2020c, Meszaros2020}, with only a handful of exceptions, such as the binary hypothesis \citep[see e.g.][]{Fernandez-Trincado2019b}. There are also a handful of known early asymptotic giant branch (early-AGB) stars in GCs \citep[see, e.g.][]{Meszaros2020}, which exhibit a nitrogen enrichment similar to that of the N-rich stars, but with a modest enrichment in carbon ([C/Fe]$>+0.15$. 

Recent studies \citep{Bekki2019, Fernandez-Trincado2020, Fernandez-Trincado2020a, Fernandez-Trincado2020b} have found that a significant fraction ($\sim66$\%) of the known metal-poor N-rich stars in the Galactic field currently reside in the inner MW. However, the site and process for the formation of these stars remain as long-standing questions \citep{Bekki2019}. The metal-poor N-rich stars stand out among the variety of metal-poor stars in our own MW, as they are uniquely suited to the application of chemical tagging, with direct bearing on the origin of the in situ structure of the inner halo and bulge. As such, these stars play a crucial role in reconstructing the formation and evolutionary history of the MW. 

NGC~6723 is an old \citep[$\gtrsim$12.50 Gyr;][]{Dotter2010, Oliveira2020}, massive \citep[1.72$\times10^{5}$ M$_{\odot}$:][]{Baumgardt2019} GC situated in the bulge region ($d_{\odot}\sim$8.3 kpc), with an orbit constraining it to lie $\lesssim2.85$ kpc from the Galactic Centre \citep{Baumgardt2019}. It is currently located at a position where the foreground interstellar reddening is low, E(B-V) $\sim0.063$ \citep{Lee2014}, making it an excellent target to identify and study the family of N-rich stars in the bulge region of the MW, in order to probe for a linkage between GC environments and field stars with nitrogen over-abundances.

The Apache Point Observatory Galactic Evolution Experiment \citep[APOGEE-2,][]{Majewski2017} is a cornerstone mission of the Sloan Digital Sky Survey IV \citep[SDSS-IV,][]{Blanton2017}, which has been designed primarily to investigate the chemical history of the MW. APOGEE and APOGEE-2 have delivered an exquisite data set by combining high-resolution near-IR spectroscopic observations from the Sloan 2.5m telescope at Apache Point Observatory (APO) and the Ir\'en\'ee du Pont 2.5m telescope at Las Campanas Observatory (LCO): the largest and most precise census of more than 24 chemical species for hundreds of thousands of stars throughout the MW. These surveys have revealed numerous giant stars with light-element abundances that deviate from the general patterns, and are at odds with current chemical-evolution models. Their origins have remained obscure, due to limitations on the numbers of such stars and the precision of the previously available data.

In this work, we take advantage of the APOGEE-2 data set to alleviate the lack of data towards the Galactic Bulge. We report on the discovery of an extra-tidal star candidate in the vicinity of a GC (NGC~6723) toward the bulge region, with nitrogen over-abundances well above that of normal Galactic field stars, and with a modest carbon enrichment. The other properties of this star, including metallicity and kinematics, strongly suggest it was once a cluster member. This is the first demonstration of a direct link between a mildly metal-poor N-rich field star with a modest carbon enrichment and a parent GC in the bulge region of the MW.

\section{Data and sample selection}
\label{section2}

We present a spectroscopic study of the bulge GC NGC~6723, and its surrounding regions, based on high-resolution ($R\sim$ 22,000) near-infrared \textit{H}-band ($\lambda\sim$ 15,145\,{\AA} to 16,960\,{\AA}, vacuum wavelengths) spectra taken as part of the APOGEE-2 survey \citep{Majewski2017, Zasowski2017}, included within the sixteenth SDSS-IV \citep{Blanton2017} data release \citep[DR16,][]{Ahumada2019}. The observations were taken at the twin APOGEE-2 spectrographs mounted on the 2.5m Sloan Foundation telescope \citep{Gunn2006} at the Apache Point Observatory in New Mexico, and on the 2.5m Ir\'en\'ee du Pont telescope \citep{Bowen1973} at Las Campanas Observatory in Chile. We refer the reader to the APOGEE-2 technical papers for further details regarding the data reduction pipeline for APOGEE-2 \citep[see, e.g.][]{Nidever2015, Garcia2016, Holtzman2018} .

We have analysed the available APOGEE-2 spectra towards the bulge GC NGC~6723, included in the APOGEE-2 fields \texttt{000-17-C} and \texttt{359-17-C}, which contain reliable spectral information for 526 stars. Probable cluster members were selected based on the nominal radial velocity (RV) of the cluster (with a velocity difference no larger than 15 km s$^{-1}$), and metallicity within 0.25 dex from the value reported in \citet[][2010 edition]{Harris1996}. The initial search was made within the blue box highlighted in panel (a) of Figure \ref{Figure1}, and limited to spectra with a SNR $>$ 60 pixel$^{-1}$. 

We decided to adopt the uncalibrated \texttt{ASPCAP} [M/H] scale, which tracks all metals relative to the Sun (listed in Table \ref{table1}), as a first guess to the stellar metallicity. This gives the overall metallicity of the stars, as it is derived by fitting the entire wavelength region covered by the APOGEE-2 spectrographs. Finally, we use Fe I lines to measure the [Fe/H].

The final sample contains eight potential members of NGC~6723. We find that seven of these stars are located inside the tidal radius, as shown in panel (b) of Figure \ref{Figure1}. There is one star, 2M18594405$-$3651518 (referred to here as the extra-tidal star), which has a RV and proper motions (see panels (a) and (c) of Figure \ref{Figure1}), and an orbital path similar to that of the NGC~6723 members. These very probable members lie in the upper part of the prominent red giant branch (RGB) in the extinction-corrected 2MASS and V$_{\rm o}$ versus (B-V)$_{\rm o}$ Colour-Magnitude Diagrams (CMDs), as displayed in panels (d) and (e) of Figure \ref{Figure1}, and have \textit{Gaia} proper motions similar to the nominal value for the cluster. 

\begin{figure*}	
	\begin{center}
		\includegraphics[width=190mm]{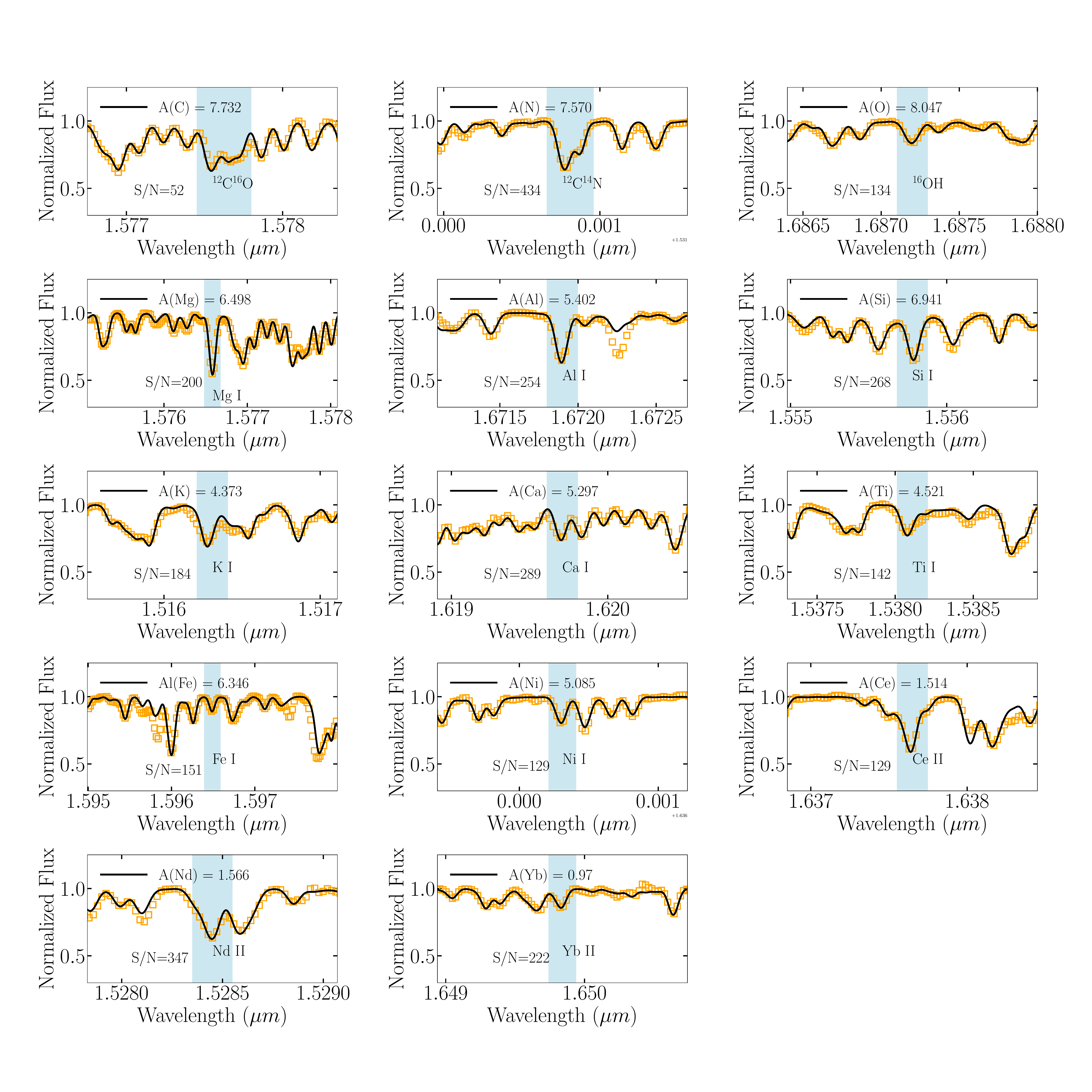}
		\caption{High-resolution near-IR \textit{H}-band spectrum of the extra-tidal star. The spectral regions are shown with orange squares. Superimposed is the best-fit of a MARCS/\texttt{BACCHUS} spectral synthesis (black line). The light blue shaded regions show the strength of the molecules ($^{12}$C$^{16}$O, $^{12}$C$^{14}$N, $^{16}$OH), and the atomic lines, namely the $\alpha$-elements (Mg I, Si I, Ca I, Ti I),  the odd-Z elements (Al I, K I), the iron-peak elements (Fe I, Ni I), and the \textit{s}-process elements (Ce II, ND II, Yb II), expressed in air wavelengths. The legends in each panel show the absolute abundance, $A$(X), of the species under consideration, and the signal-to-noise (S/N) in the regions of the features, respectively.}
		\label{Figure2}
	\end{center}
\end{figure*}

Following the same methodology and techniques described in several APOGEE papers \citep{Fernandez-Trincado2016a, Hawkins2016, Fernandez-Trincado2017, Fernandez-Trincado2019a, Fernandez-Trincado2019b, Fernandez-Trincado2019c, Fernandez-Trincado2019d, Fernandez-Trincado2020c}, we employed the \texttt{BACCHUS} software \citep{Masseron2016} to manually analyze each star of our sample, in order to re-examine the reliability of each atomic and molecular line present in each spectrum, and provide chemical abundances for 14 chemical species, listed in Table \ref{table1}. These abundances have been computed adopting a line-by-line approach under the assumption of local thermodynamic equilibrium (LTE). With this independent methodology, we obtain complementary abundance ratios not provided by the \texttt{ASPCAP} pipeline \citep{Garcia2016}, in particular for the heavy neutron-capture (\textit{s}-process) elements (Ce II, Nd II, and Yb II). Figure \ref{Figure2} illustrates the best spectral-synthesis calculation on clean selected features for the extra-tidal star. 

Importantly, none of the newly identified stars in this work have strong RV variability over the temporal span of the APOGEE-2 observations (visit-to-visit variations, $\sigma RV < 1$  km s$^{-1}$), which were obtained during 2018-05-24 (visit 1) and 2018-05-25 (visit 2). Therefore, with the current data there is no evidence that any of these objects are members of binary systems. 

Furthermore, we found that our sample has a \textit{Gaia} re-normalised unit weight error (\texttt{RUWE}) value $<$ 1.4, as listed in Table \ref{table1}, indicating that they are astrometrically well-behaved sources in the \textit{Gaia} Early Data Release 3 (\textit{Gaia} EDR3) catalogue \citep{Brown2020}. 

Panels (a) to (c) of Figure \ref{Figure1} show the metallicity versus kinematic, sky positions, and astrometric properties of our sample, as well as the serendipitous discovery of an chemically atypical extra-tidal star in the vicinity of NGC~6723; a result not previously seen. We find that 5 of the 7 cluster members are strongly enhanced in nitrogen, [N/Fe]$>+0.86$, well above the Galactic levels (which is typically $\lesssim+0.5$), and a clear indicator of the presence of nitrogen-enhanced stars in NGC~6723. The CMDs displayed in Figures \ref{Figure1} (d) and (e) show that most of the stars in our sample lie in the upper part of the prominent red giant-branch (RGB) of NGC~6723, and do not exhibit a wide metallicity variation. This is also supported by the spectroscopic values.

\section{Atmospheric parameters and elemental abundances}
\label{section3}

\begin{figure*}	
	\begin{center}
		\includegraphics[width=90mm]{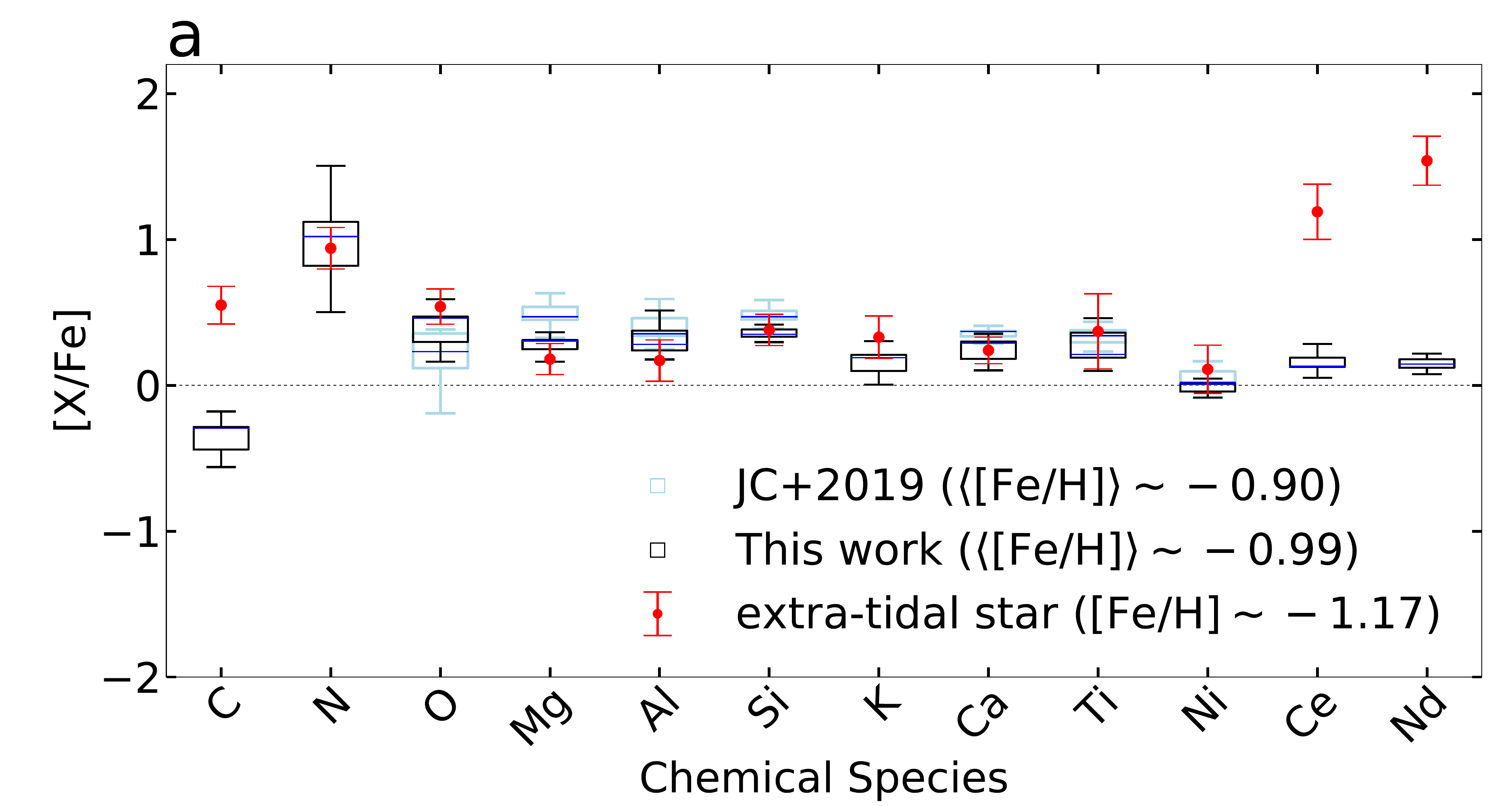}\includegraphics[width=90mm]{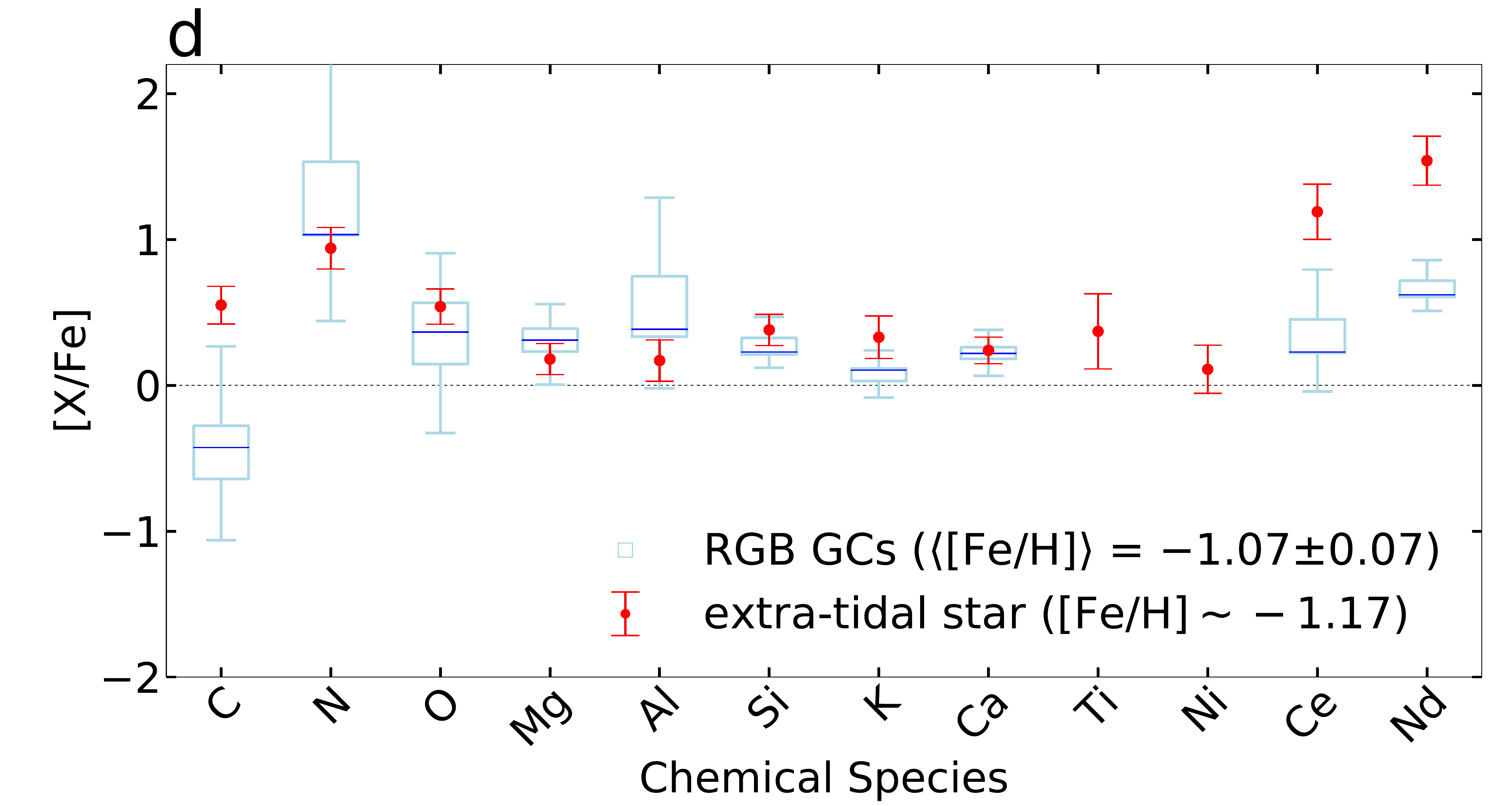}
		\includegraphics[width=90mm]{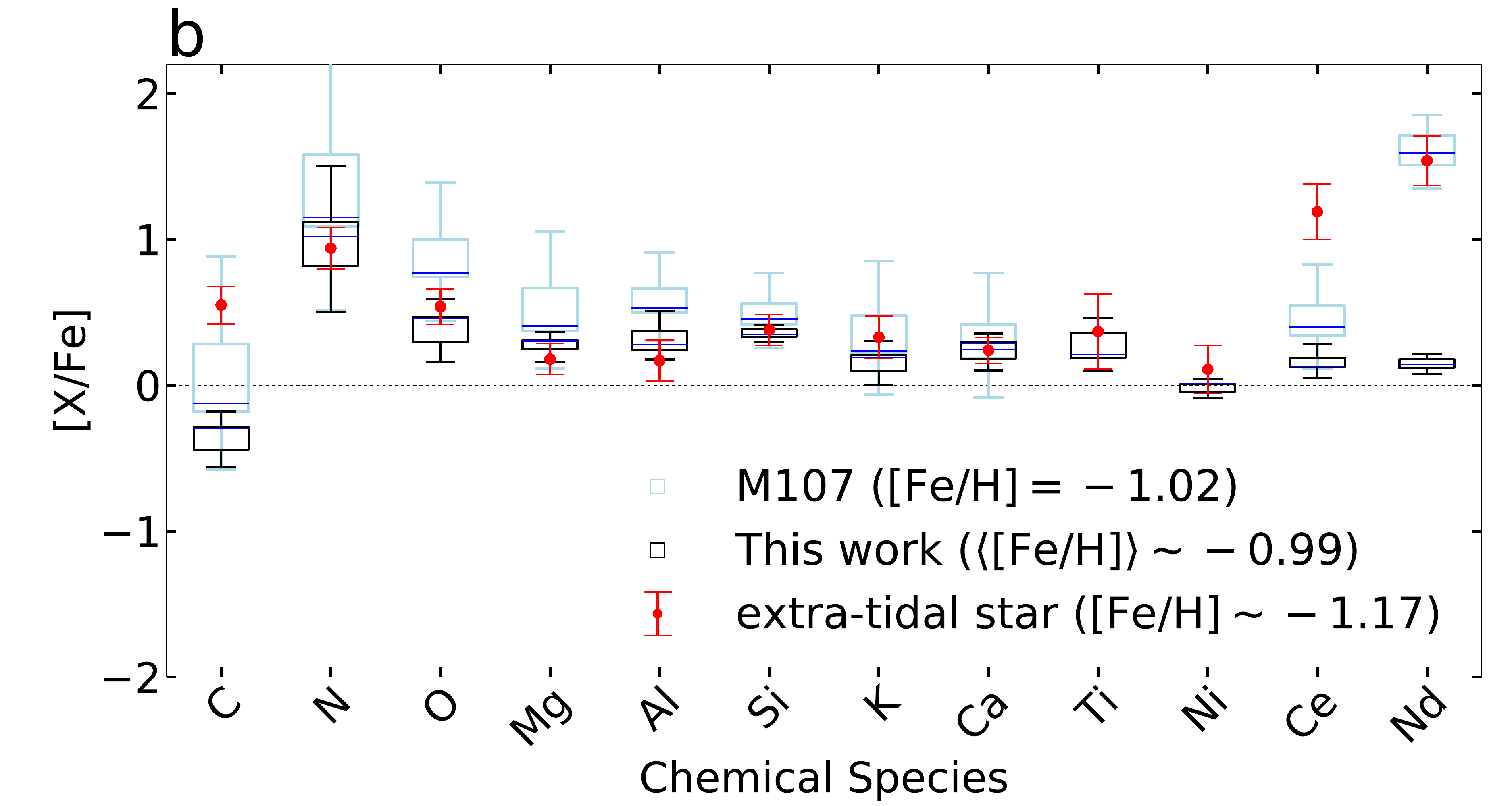}\includegraphics[width=90mm]{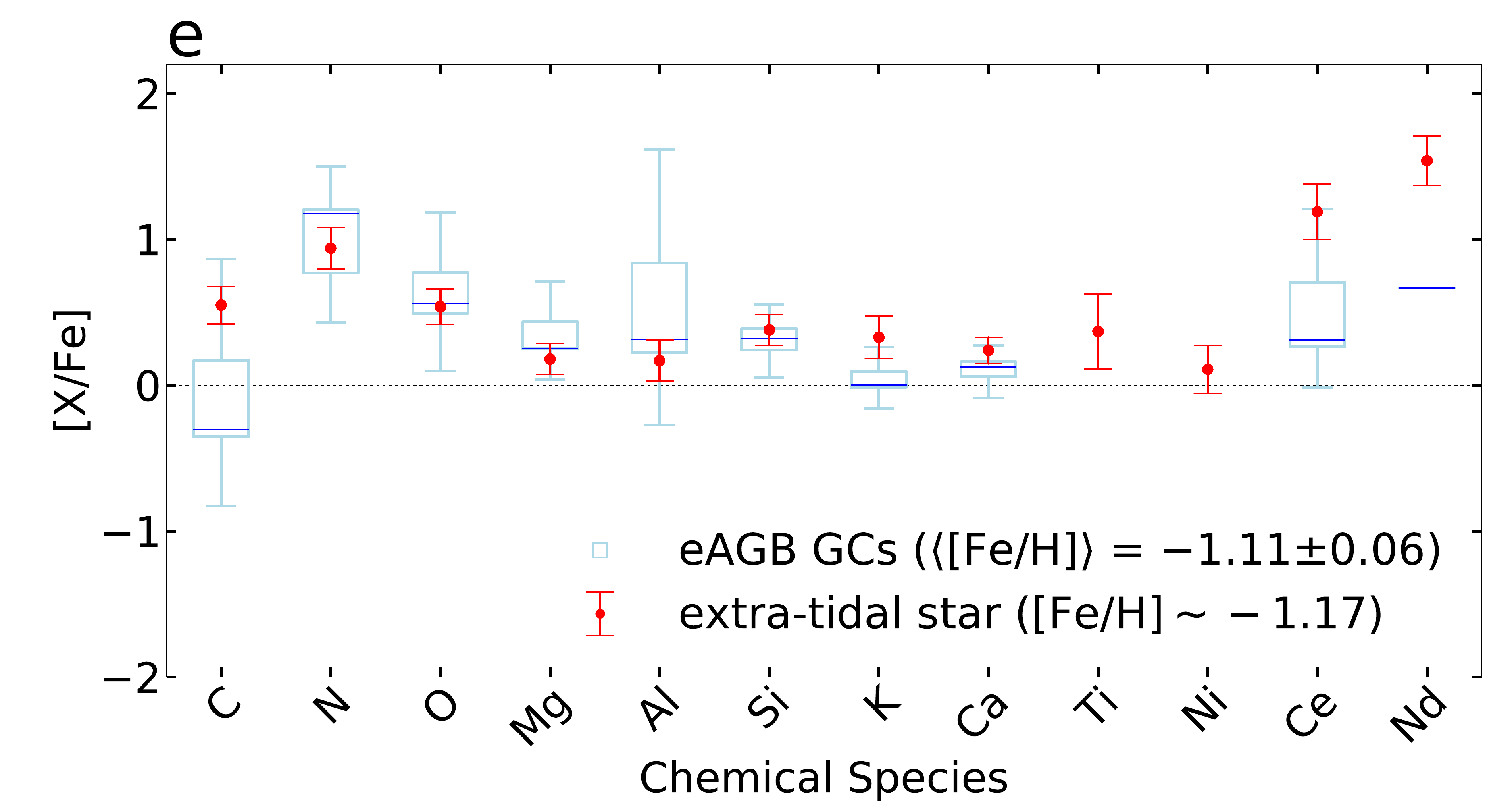}
		\includegraphics[width=90mm]{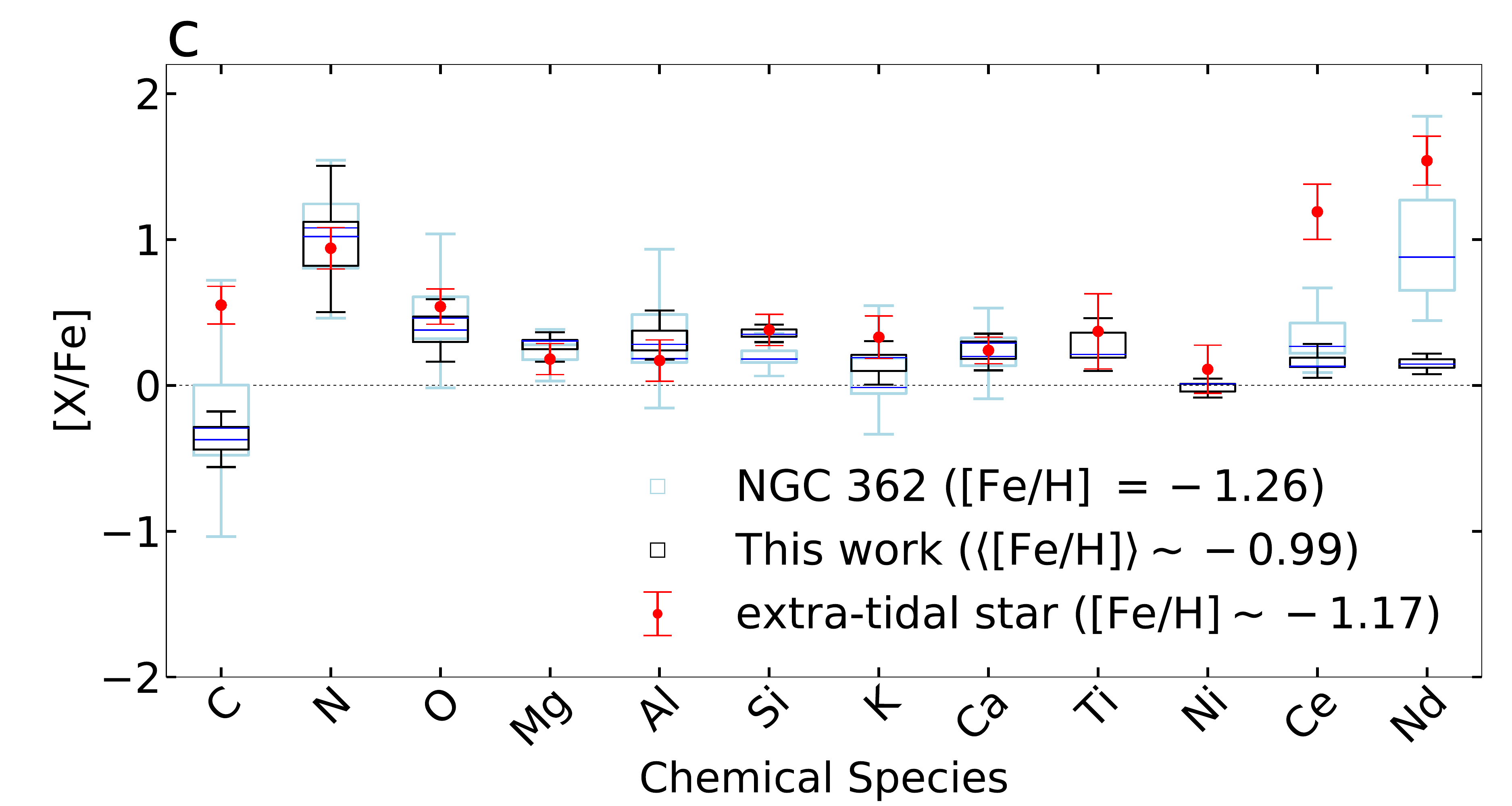}\includegraphics[width=90mm]{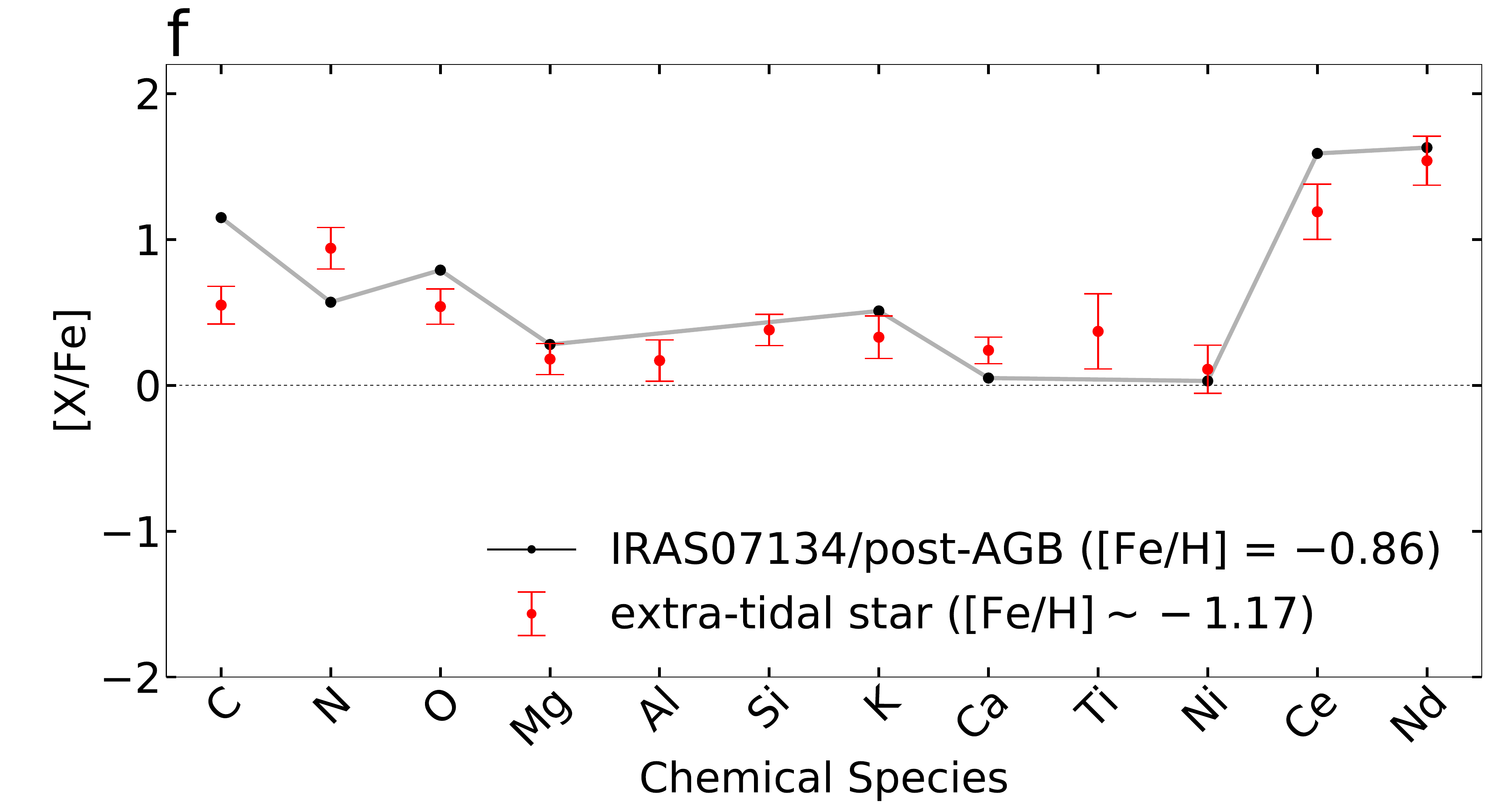}
		\caption{Comparison of elemental-abundance ratios of likely member stars for NGC~6723 with other GCs of similar metallicity, and the extra-tidal star with RGB and early-AGB (eAGB) cluster stars, and IRAS07134 star. Panel (a): Elemental-abundance patterns for our sample (black boxes and red dot) compared to the spectroscopic study of NGC~6723 \citep{Crestani2019} (light blue boxplots). Panel (b):  Comparison with M 107 \citep{Meszaros2020}. Panel (c): Comparison with NGC~362 \citep{Meszaros2020}. Panel (d): Elemental-abundance patterns for the extra-tidal star compared to those of RGB GC stars. Panel (e): Compared with early-AGB GC stars \citep{Meszaros2020}. Panel (f): Comparison with a field post-AGB star, IRAS 07134 \citep{DeSmedt2016}, with comparable atmospheric parameters. Boxes represent the inter-quartile ranges (IQR), whiskers the 1.5$\times$IQR limits, and blue lines the medians.}
		\label{Figure3}
	\end{center}
\end{figure*}	

We used the \texttt{BACCHUS} code \citep{Masseron2016} to derive the metallicity (from Fe I lines), broadening parameters, and chemical abundances for the stars in our sample, based on careful line selection, and carried out a detailed inspection of each APOGEE-2 spectrum in order to examine the reliability of the abundance ratios, in the same manner as described in \citet{Fernandez-Trincado2019c}. The adopted atmospheric parameters and  the typical uncertainties are listed in Table \ref{table1}.

The \texttt{BACCHUS} code relies on the radiative transfer code \texttt{Turbospectrum} \citep{Alvarez1998, Plez2012} and the \texttt{MARCS} model atmosphere grid \citep{Gustafsson2008}. For each element and each line, the abundance determination proceeds as in previous APOGEE-2 works \citep{Hawkins2016}. In summary, the steps are: (\textit{i}) a spectrum synthesis, using the full set of (atomic and molecular) lines to find the local continuum level via a linear fit; (\textit{ii}) cosmic and telluric line rejections are performed; (\textit{iii}) the local signal-to-noise ratio (S/N) per element is estimated; (\textit{iv}) a series of flux points contributing to a given absorption line are automatically selected; and (\textit{v}) abundances are then derived by comparing the observed spectrum with a set of convolved synthetic spectra characterised by different abundances.

Four different abundance determinations are used: (\textit{i}) line-profile fitting; (\textit{ii}) core line-intensity comparison; (\textit{iii}) global goodness-of-fit estimate; and (\textit{iv}) equivalent-width comparison. Each diagnostic yields validation flags. Based on these flags, a decision tree then rejects or accepts the line, keeping the best-fit abundance. We adopted the $\chi^{2}$ diagnostic as the abundance determinant, because it is considered to be the most robust. However, we stored the information from the other diagnostics, including the standard deviation between all four methods. The line list used in this work is the latest internal DR14 atomic/molecular linelist (\texttt{linelist.20170418}), including the $s$-process elements (Ce II, Nd II, and Yb II)--\citep{Hasselquist2016, Cunha2017}. For a more detailed description of these lines, we refer the reader to a forthcoming paper (Holtzman et al. in preparation).

In particular, a mix of heavily CN-cycled and $\alpha$-poor \texttt{MARCS} models were used, as well as the same molecular lines adopted by APOGEE-2 \citep{Smith2013}, in order to determine the C, N, and O abundances.  In addition, we have adopted the C, N, and O abundances that satisfy the fitting of all molecular lines consistently; that is to say, we first derive $^{16}$O abundances from $^{16}$OH lines, then derive $^{12}$C from $^{12}$C$^{16}$O lines, and $^{14}$N from $^{12}$C$^{14}$N lines; the C--N--O abundances were derived iteratively to minimize the $^{16}$OH, $^{12}$C$^{16}$O, and $^{12}$C$^{14}$N dependences \citep{Smith2013}.

It is important to perform consistent chemical-abundance analyses using atmospheric parameters determined independently, in order to check for any significant deviation in the derived abundances. To achieve this, the photometric effective temperatures were calculated using the $J-K_{s}$ (2MASS) colour-relation methodology \citep{Gonzalez2009}. For the extra-tidal star, we adopted the extinction correction obtained from the Rayleigh Jeans Colour Excess (RJCE) method \citep{Majewski2011}. Thus, the photometry is extinction corrected, adopting E(B-V)$=$ 0.063 \citep[see, e.g.,][]{Lee2014} for stars inside the cluster tidal radius, and assuming an E(B-V) = 0.520 \citep{Majewski2011} for the extra-tidal star, which lies at a region outside the cluster that is severely affected by extinction, given their position toward the Galactic bulge where the reddening variation may be substantial \citep[see, e.g.][]{Alonso-Garcia2017}, even if the projected distance between the extra-tidal star and cluster is small.

We assumed a surface gravity from the \texttt{PARSEC} isochrones \citep{Bressan2012}, using $\approx$12.5 Gyr as the estimated age of NGC~6723 \citep[][]{Dotter2010, Oliveira2020}, and the uncalibrated metallicity ([M/H]), as derived by \texttt{ASPCAP}/APOGEE-2 runs. The adopted stellar parameters are listed in Table \ref{table1}.

The adoption of a purely photometric temperature scale enables us to be somewhat independent of the \texttt{ASPCAP}/APOGEE-2 pipeline, which provides important comparison data for future pipeline validation. The final results presented in this paper are based on computations done with the \texttt{BACCHUS} code using the mentioned photometry and atmospheric parameters, as listed in Table \ref{table1}. 

\begin{figure*}	
	\begin{center}
		\includegraphics[width=185mm]{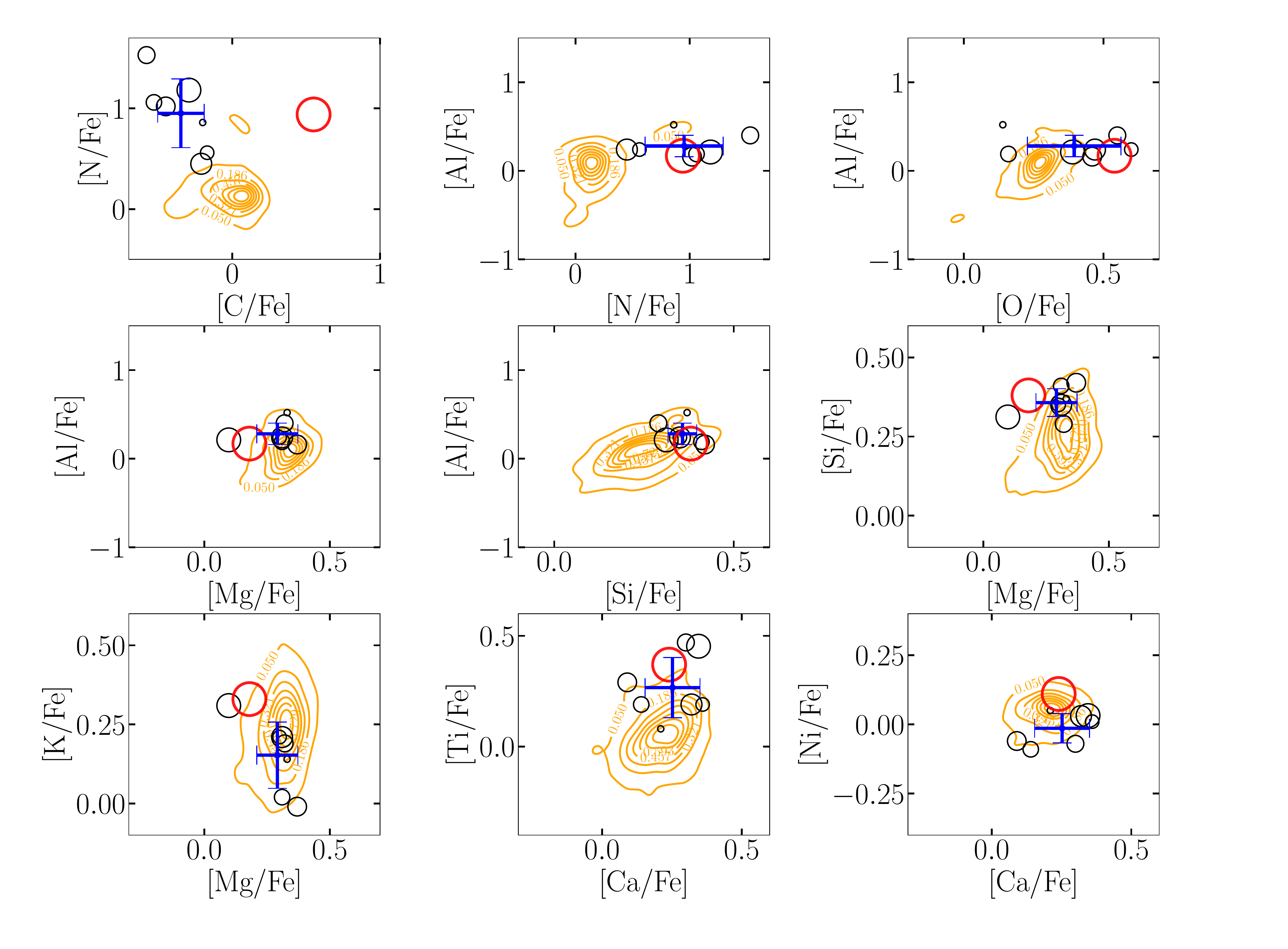}
		\caption{Distributions of various elemental-abundance ratios from the APOGEE-2 survey for bulge field stars in close proximity to the Galactic Centre ($-20^\circ<l<20^\circ$ and $|b|<20^\circ$), and comparison with stars in NGC~6723. The field-star abundances are represented by orange iso-abundance contours. The member stars of NGC~6723 and the extra-tidal star analyzed in this work are shown with black and red open circles, respectively. The circle sizes reflect their ${\rm V_o}$ magnitudes, with decreasing size for fainter magnitudes. The blue cross symbol denotes the average $\langle$[X$_{\rm i}$/Fe]$\rangle$ and its associated star-to-star scatter ($\sigma_{\rm \langle[X_i/Fe]\rangle}$) of the abundance derived from the seven member NGC~6723 stars analyzed in this work.}
		\label{Figure5}
	\end{center}
\end{figure*}	

The abundance values are sensitive to all of the atmospheric parameters, depending on the  chemical species. To estimate their uncertainties, we have varied the atmospheric parameters one at a time by the typical values of $\Delta$T$_{\rm eff} = \pm 100$ K, $\Delta \log g=\pm0.3$ dex, and $\Delta\xi_t = \pm0.05$ km s$^{-1}$, and then computed the abundances for all species for each of these possibilities for two stars -- one cluster member (2M18594898$-$3635401) and the extra-tidal star (2M18594405$-$3651518). Thus, each line for each star has a corresponding $\sigma_{\rm T_{eff}}$, which refers to its response to $\Delta {\rm T_{eff}}$; $\sigma_{\log g}$ the response to $\Delta{\log g}$; $\sigma_{\rm \xi_{t}}$ the response to $\Delta{\rm \xi_{t}}$, and the uncertainties on the mean due to line-to-line scatter. The uncertainties are propagated in quadrature to compute the uncertainty of each chemical species in the two stars. The computed uncertainties are listed in Table \ref{table2}.

Even though the \texttt{BACCHUS} code has its own procedure to include or reject lines on a star-by-star basis, it is still important to select the lines beforehand, due to the uncertainty related to the synthesis approach, such as line saturation. All the selected atomic and molecular lines were visually inspected to ensure that the spectral fit was adequate. 

The individual oxygen abundances are also listed in Table \ref{table1}. We caution about the accuracy of [O/Fe], as they display larger scatter, which may be due to telluric features and uncertain determinations in the T$_{\rm eff}$ regime of our objects. As highlighted by previous works, the uncertainty arises because \texttt{BACCHUS} determines these abundances from the strengths of $^{12}$C$^{14}$N and $^{12}$C$^{16}$O lines, which become too weak for stars at relatively low metallicities ([Fe/H]$\lesssim -1.0$). Figures \ref{Figure2} and \ref{Figure3} clearly shows that K I, Ce II, Nd II, and Yb II enhancement of the extra-tidal star compared to other NGC 6723 stars can be convincingly claimed. Consequently, we conclude that the Ce, Nd, and Yb enhancement was likely inherited from the initial gas composition of NGC~6723.

\section{Results}
\label{section4}

\subsection{NGC~6723 stars}

\begin{figure*}
	\centering
	\includegraphics[width=180mm]{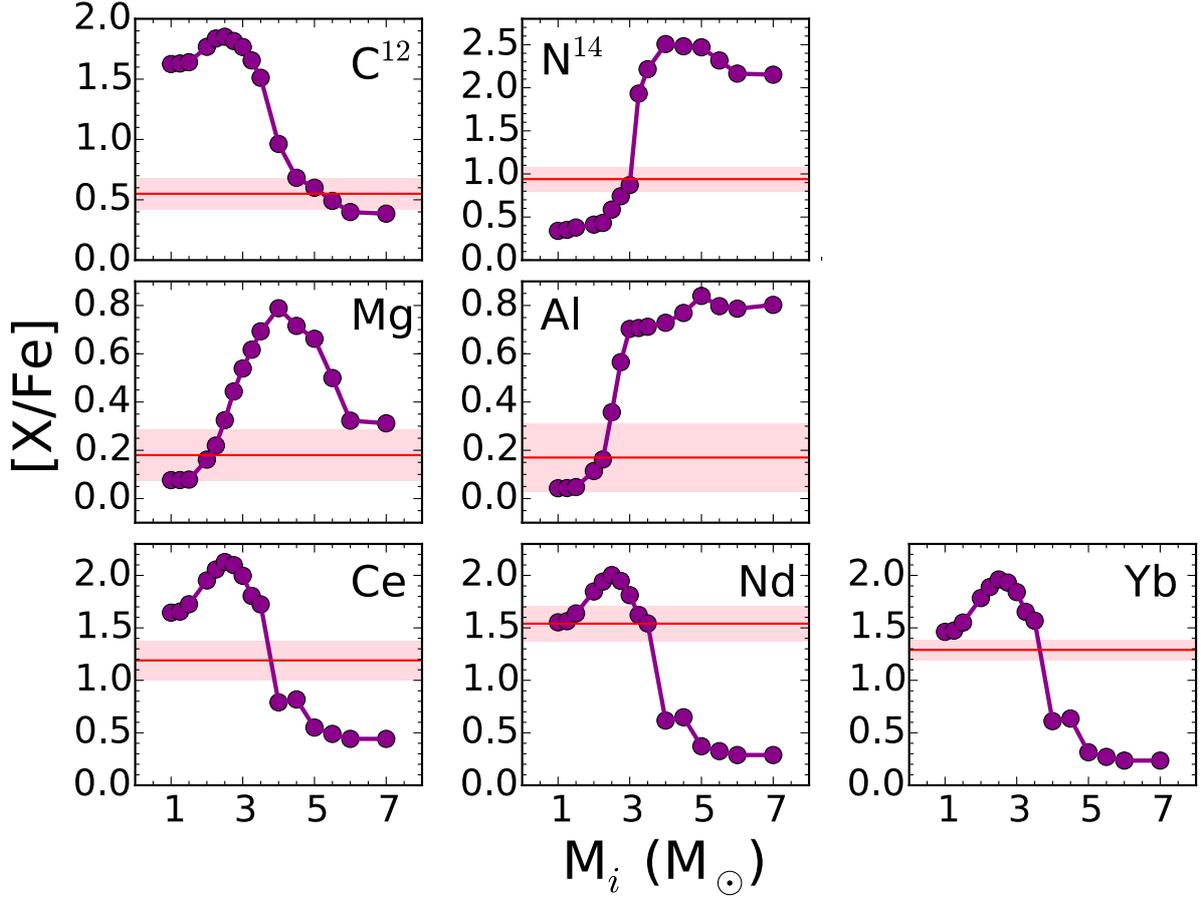}
	\caption{Predicted abundance ratios, as a function of the initial stellar mass, for low-and-intermediate-mass stars. ${\rm [X_{i}/Fe]}$ values were obtained from stellar yields by \citet{Karakas2014, Karakas2018}. For comparison, the red line and shaded red bands represent the observed abundance ratios and sensitivities for our extra-tidal star.}  
	\label{Figure4}
\end{figure*}	

For the selected stars, we derive abundance ratios covering the main element families, namely the light elements (C, N), the $\alpha$-elements (O, Mg, Si, Ca, Ti),  the odd-Z elements (Al, K), the iron-peak elements (Fe, Ni), and products of the slow neutron-capture (\textit{s}-) process (Ce, Nd, Yb), and have abundance determinations from spectra in the \textit{H}-band of APOGEE-2 \citep{Meszaros2020}. For the cluster members, we find metallicity consistent with previous estimates for this cluster, $\langle$[Fe/H]$\rangle$ $= -1.00$, with a small, 0.06 dex, star-to-star scatter, commensurate with recent results based on high-resolution spectroscopy \citep{Rojas-Arriagada2016, Crestani2019}, which indicate a mono-[Fe/H] behaviour. 

We obtained a mean nitrogen abundance ratio of [N/Fe] $= +0.95$, with an observed star-to-star scatter ($0.34$ dex) that well exceeds the observational uncertainties. The abundance ratios for the other elements up to the iron peak (O, Mg, Al, Si, K, Ca, Ti, and Ni) appear to be homogeneous within the permitted error variation, and replicate the chemical patterns observed for Galactic GCs of comparable metallicity, as shown in panels (b) and (c) of Figure \ref{Figure3}. The heavy \textit{s}-process  elements, such as Ce and Nd, are lower ([Ce/Fe], [Nd/Fe] $\lesssim +0.29$) compared to M107, a GC with similar metallicity as NGC~6723, excepting Yb, which we find to be higher in some NGC~6723 stars. It has been noted previously that \textit{s}-process-element enhancements $\gtrsim +0.4$ are usually rare, and are reported for only a few clusters \citep{Meszaros2020}, particularly at the metallicity of NGC~6723, suggesting they arise from pollution by AGB stars \citep{Ventura2016, Meszaros2020}. 

The large star-to-star abundance variation inside the cluster -- especially in the case of N -- is indicative of the presence of multiple populations (MPs) in NGC~6723, as found in other massive ($\gtrsim10^{5}$M$_{\odot}$) GCs at all metallicities. There is thus strong evidence for an intrinsic spread in [N/Fe], including a clear C--N anti-correlation. However, we find no clear Al--N correlation or Al--O anti-correlation (see Figure \ref{Figure5}). In other words, NGC~6723 exhibits MPs based on the N abundances, despite appearing to have single populations in the Al abundances. This is similar to other Galactic GCs at this metallicity \citep{Meszaros2020}, for which it has been suggested that lower spreads in Al could be ascribed to operation of a modest Mg-Al cycle. If the intra-cluster polluters were in fact low-mass ($\lesssim 3$ M$_{\odot}$) stars, we would expect low Al production \citep{Meszaros2020} in NGC~6723. The low mean abundance and small Al spread ($\langle$ [Al/Fe]$\rangle = +0.28\pm0.12$) measured for the NGC~6723 sample supports this conclusion.

The stars we have studied in NGC~6723 have [C/Fe]$< -0.17$, and display a similar C--N anti-correlation as M107, with $\langle$[C/Fe]$\rangle= -0.35\pm0.16$. Additionally, NGC~6723 exhibits a star-to-star [Mg/Fe] scatter with no significant [Al/Fe] spread. No Mg-Al anti-correlation is apparent, and the scatter is small. In addition, inspection of the [Si/Fe] ratio, as a function of [Al/Fe] or [Mg/Fe], for the few stars with reliable Si measurements, did not reveal any clear trend. This suggests no net production of these elements, but rather is the likely result of the conversion of Mg into Al during the Mg-Al cycle \citep{Meszaros2020}.

Figure \ref{Figure5} compares the C, N, O, Mg, Al, Si, Ca, Ti, Ni, and Ce abundances of our sample with Galactic Bulge stars. The behaviour of those chemical species matches the mean [X$_{\rm i}$/Fe] (with X$_{\rm i}$ meaning the chemical species) of MW stars at similar metallicity, while the [N/Fe] abundance ratios are clearly super-Solar. Our results for [Si/Fe] are in reasonable agreement with optical observations \citep{Rojas-Arriagada2016, Crestani2019}.

NGC~6723 has a uniform and constant [Ca/Fe] abundance ratio ($\langle $[Ca/Fe]$\rangle = +0.25 \pm 0.10$), in agreement with optical observations \citep{Crestani2019}, which indicates that it is not affected by the H-burning process \citep{Meszaros2020}, as Ca is mostly produced by supernovae. For the odd-Z element K, we find that NGC~6723 exhibits a weak K--Mg anti-correlation, with $\langle$[K/Fe]$\rangle = +0.15 \pm 0.10$, suggesting that this population might have formed from super-AGB ejecta \citep{Ventura2012}. 

For the remaining chemical species, we find that the average $\langle$[O/Fe]$\rangle$ ($+$0.40 $\pm$0.17), $\langle$[Ti/Fe]$\rangle$ ($+0.27\pm0.14$), $\langle$[Ni/Fe]$\rangle$ ($-$0.01$\pm$0.05), $\langle$[Ce/Fe]$\rangle$ ($+0.14\pm0.08$), $\langle$[Nd/Fe]$\rangle$ ($+0.15\pm0.07$), and  $\langle$[Yb/Fe]$\rangle$ ($+0.68\pm0.09$) ratios are comparable to values for M107 and NGC~362, and in agreement with previous results \citep{Rojas-Arriagada2016, Crestani2019}. However, some differences are noteworthy for Nd, compared to GCs at similar metallicity, but it does agree with the production of other $s$-process species such as Ba \citep{Rojas-Arriagada2016}, with small star-to-star scatter.

\begin{figure*}
	\centering
	\includegraphics[width=185mm]{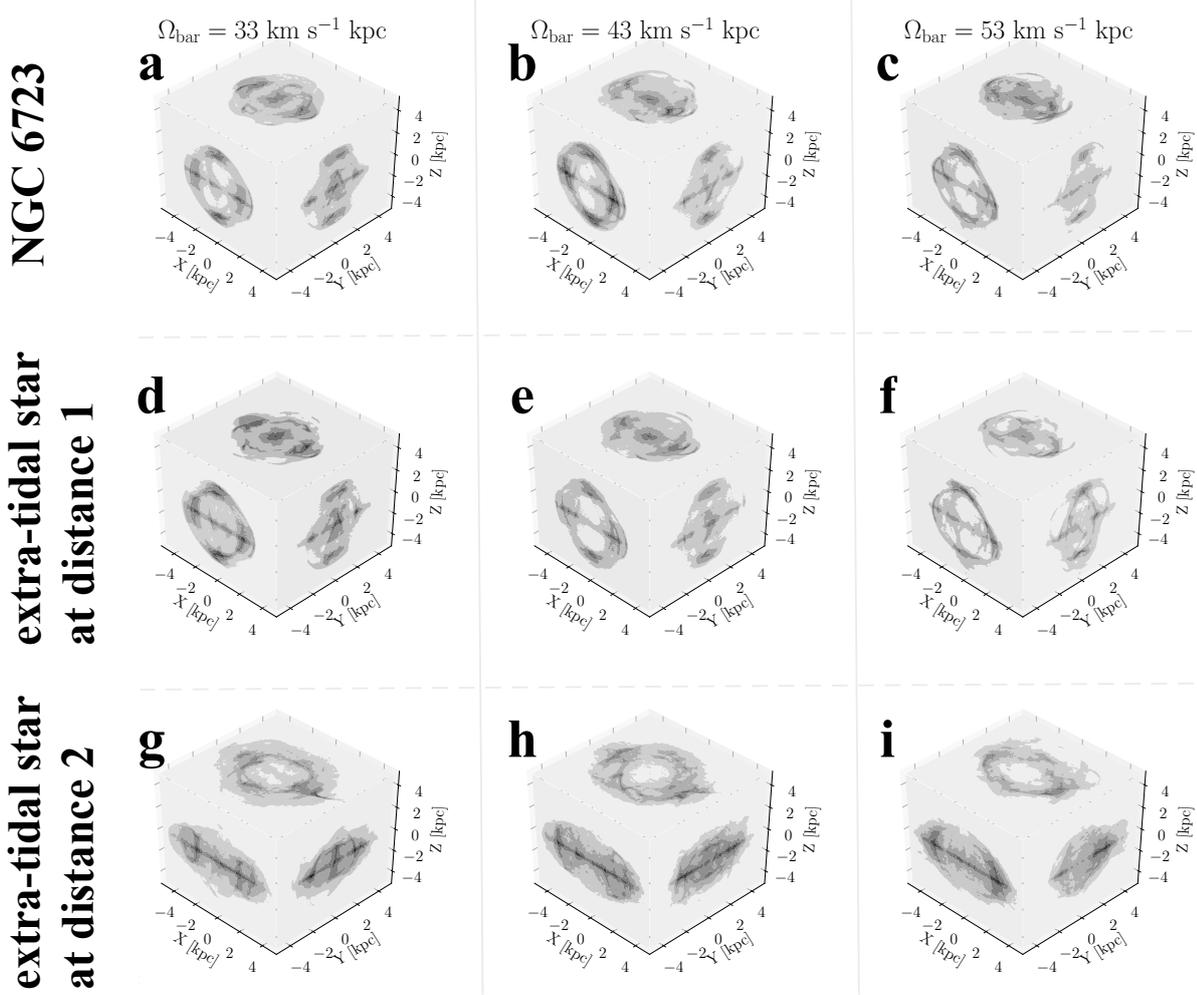}
	\caption{Thousand-orbit realisations of NGC~6723 and the extra-tidal star time-integrated for 2 Gyr. The dark colours correspond to the most probable regions of the space, which are crossed more frequently by the simulated orbits, assuming three different values of the angular velocity of the bar ($\Omega_{\rm bar} = 33, 43,$ and 53 km s$^{-1}$ kpc). Panels (a), (b), and (c) show the orbits of NGC~6723, using as initial conditions the observed values from \citet{Baumgardt2019}; the middle and bottom rows show the orbits of the extra-tidal star, adopting the observed values from the Table \ref{table1}, and an assumed heliocentric distance at 8.3 kpc (panels d, e, and f), and 6.24 kpc (panels g, h, and i).}  
	\label{Figure8}
\end{figure*}	

\subsection{Extra-tidal star candidate}

Panels (d) and (e) of Figure \ref{Figure1} show the existence of an extra-tidal star in the smoothly curved morphology of the upper RGB in the V$_{\rm o}$ versus (B-V)$_{\rm o}$ \citep{Lee2014}, and H$_{\rm o}$ versus (J-K$_{\rm s}$)$_{\rm o}$ diagrams, extending from the cluster centre out to $\sim$1.3 tidal radii. The brightness and bluish offset in all three diagrams from the equally bright RGB stars indicate that this extra-tidal star is in reasonable agreement with the expected behaviour for AGB stars in GCs. Figure \ref{Figure1} (a) also shows that this extra-tidal star has similar metallicity and radial velocity, and Figures \ref{Figure1} (b) and (c) demonstrate the similarity of the orbital paths and proper motions to NGC~6723 member stars as well. Thus, these properties are vetted not only by chemical abundances, but by photometry and kinematics as well. 

The $\alpha-$element (Mg, Si, Ca, and Ti), the odd-Z element (Al and K), and the iron-peak element Ni abundance ratios fall between $-0.09$ and $+0.52$ for our extra-tidal star. This is similar to the members of NGC~6723 and other GC stars with comparable metallicity, within the uncertainties (see Table \ref{table2} and Figure \ref{Figure3}).  Figure \ref{Figure3} also shows that the extra-tidal star is strongly enhanced in the \textit{s}-process elements (Ce II, Nd II, and Yb II), which is in agreement with other GC stars at similar metallicity \citep{Meszaros2020}. \\

Figure~\ref{Figure2} reveal that the newly identified extra-tidal star has a stellar atmosphere strongly enriched in $^{12}$C$^{14}$N features, which indicates a high enrichment in nitrogen ([N/Fe]$=+0.94$), well above usual Galactic levels. However, this particular star has a carbon abundance higher than the abundance ratios measured for NGC~6723 member stars as shown in panel (a) of Figure~\ref{Figure3}, but in reasonable agreement with the most C-rich stars in GCs of comparable metallicity (see panels (b) and (c) of Figure~\ref{Figure3}). This atypical chemical pattern suggests that the extra-tidal star is not a genuine second-generation star as other typical N-rich field /GC stars, but indeed is part of a second-type of N-rich star with modest carbon enrichment, likely a GC star with AGB-like chemical patterns. Figures~\ref{Figure2} and \ref{Figure3} also clearly show that K I, and Ce II enhancement of the extra-tidal star can be convincingly claimed. Thus, we conclude that the Ce, Nd, and Yb enhancement was inherited from the initial gas composition of NGC~6723. 

We also find that the oxygen abundance is in reasonable agreement with levels seen for NGC~6723 stars \citep{Rojas-Arriagada2016, Crestani2019}. From the strength of the $^{12}$C$^{16}$O and $^{16}$OH molecular features (see Figure~\ref{Figure2}), one would expect that more oxygen in the atmosphere means that more carbon is locked into $^{12}$C$^{16}$O, and is not available to form $^{12}$C$^{14}$N, as the strength of the $^{12}$C$^{14}$N features is anti-correlated with [O/Fe]. From Table \ref{table1}, even a small change in the surface temperature and $\log g$ requires some adjustment to the carbon and oxygen abundance, but [N/Fe] is relatively insensitive to small T$_{\rm eff}$ and surface gravity changes, and we can be confident that this star is indeed nitrogen enriched.

\section{Possible origins for the observed chemical composition of the extra-tidal star}
\label{section5}

There are several competing scenarios that might explain the unusual chemical abundances of our extra-tidal star, in the context of an escaped member of NGC~6723 that was tidally disrupted and captured by the MW's bulge. 

It appears likely that some stars (C-poor or C-enriched) belonging to the family of mildly metal-poor N-rich stars (peaking at [Fe/H] $\sim-1.0$) could be relics of initially massive clusters such as NGC~6723. It is reasonable to assume that some of those objects were formed in Galactic GCs and later dynamically ejected into the Galactic field when their parent GCs were ultimately disrupted and destroyed, possibly during disc--bulge crossing events \citep[see, e.g.,][]{Leon2000, Lane2010,  Fernandez-Trincado2015a, Fernandez-Trincado2015b, Fernandez-Trincado2016b, Kundu2019a, Kundu2019b, Kundu2020, Hanke2020, Sollima2020}. However, this has not heretofore been directly demonstrated.

With the newly analyzed extra-tidal star, we want to investigate the possible origin of neutron-capture (\textit{s}-process) enrichment, simultaneous with the observed enrichment in nitrogen and its apparent carbon enrichment. 

Figure \ref{Figure4} shows the observed photospheric abundances of our target star, along with theoretical (C, N, Mg, Al, Ce, Nd, and Yb) abundance ratios from nucleosynthetic AGB models closest to [M/H] $\sim -1.09$. We obtained these models by interpolating the stellar yields from \citet{Karakas2014} and \citet{Karakas2018}. From inspection, an over-abundance of the elements created by the \textit{s}-process (Ce II, Nd II, and Yb II) support the idea that AGB stars with initial masses of $\sim$3.7 M$_{\odot}$ could reasonably explain the observed \textit{s}-process enrichment. As also seen in this figure, the C and N enrichments could be explained, at $\sim$2$\sigma$, with initial masses of $\sim$3.7 and $\sim$2.5 M$_{\odot}$, respectively, while Mg and Al suggest an initial mass of $\sim$2.5 M$_{\odot}$. Thus, most of these species are in agreement with production by an AGB star having an initial mass $\sim$2.5 or 3.7 M$_{\odot}$, which underwent a mass-transfer event, resulting in pollution of an intermediate-mass companion that today has evolved to the tip of the RGB. We caution about the accuracy of the estimated [O/Fe] values, as they display a larger scatter, possibly due to telluric features and uncertain determinations in the T$_{\rm eff}$ regime of our objects. As highlighted in previous works, the uncertainty arises because \texttt{BACCHUS} determines these abundances from the strengths of $^{12}$C$^{14}$N and $^{12}$C$^{16}$O lines, which become too weak for stars at relatively low metallicities ([Fe/H] $-1.0$). For this reason, we do not draw conclusive constraints in the mass of the progenitor based on the individual oxygen abundance listed in Table \ref{table1}.

It is  also possible that this star is itself an AGB star (perhaps with no relationship to NGC~6723), which could also explain the puzzle of its high nitrogen and \textit{s}-process enrichment. However, such a range in mass would correspond to a very young star ($\sim$0.5 -- 0.7 Gyr), which is likely too low to support this hypothesis. The AGB stage of stellar evolution is also very short-lived, so there is a low probability of finding an AGB star. 

Taking into account the above results, we speculate on several possible scenarios that could be viable to explain the unusual nature of our object, in the context of an escaped member of NGC~6723 that was tidally disrupted and captured by the bulge of the MW. 

\begin{itemize}
	\item[i.]  \textit{Extrinsic mechanism} (binary-mass transfer system). The over-abundance of \textit{s}-process elements could come from the accretion of \textit{s}-process-rich matter from a former thermally pulsing (TP)-AGB companion during its heavy mass-loss phase on the AGB \citep{Brown1990}, which has since evolved into a faint white dwarf \citep[see,][for instance]{Fernandez-Trincado2019b}. Figure \ref{Figure4} indicates that our measured abundance ratios, from the light elements (C, N), the $\alpha$-element (Mg), and the odd-Z element (Al), to the $s$-process elements (Ce, Nd, and Yb), could have  originated from material previously enriched in a TP-AGB companion with an initial mass of $\sim 2.5$ and $\sim3.7$ M$_{\odot}$, respectively. 
	
	Consequently, if the case of an extrinsic mechanism is favoured, and the extra-tidal star is/was part of a binary system before/after leaving the cluster, it would be possible that the star accreted N-rich material from the companion when the latter reached the AGB phase with the inferred mass. In this case, the abundances that we measure are not its original ones, but they reflect the chemical composition of the interior of the companion, plus some degree of dilution in the convective envelope of the accreting star. In addition to the nitrogen and \textit{s}-process enrichment, carbon should also be over-abundant, as a result of a third dredge-up episode in the donor. Such a star could have polluted the extra-tidal star with C, N, and \textit{s}-process-rich material, but lost so much mass that the binary may be disrupted, or become so wide that RV variations would be very small, which could explain the lack of a detectable white dwarf companion. However, the current PMs and radial velocity of the extra-tidal star appear to support that this object was detached from the cluster by tidal disruption, consistent with the low relative velocity between this cluster and the star during close encounters. In such a case, the possibility that our object was part of a binary system before departure from the cluster would not be supported \citep[see e.g.][]{Fernandez-Trincado2016a}.
	
	  The absence of RV variation over the temporal span of the APOGEE-2 observations ($< 1$ day) and \textit{Gaia} EDR3 ($-93.22\pm1.23$ km s$^{-1}$) provide no evidence that our object (the extra-tidal star) is currently a binary with a white dwarf companion. Long-term RV monitoring of this unusual object would naturally be the best course to rule out an origin by the binary channel. 
	
    If there is any possibility to support the binary hypothesis in the future, perhaps from careful long-term RV monitoring, then it is worth mentioning that, if the donor material was lost in a wind from a $\sim$2.5--3.7 M$_{\odot}$ AGB star, it must have been shed very recently. However, this scenario is not compatible with its apparent brightness in Figure~\ref{Figure1} (d) and (e), as the object is intrinsically bluer and more luminous than other observed stars in NGC~6723.
	
	\item[$\bullet$] Intrinsic mechanism. A second scenario is that our unusual object could itself be an AGB star.  In this case,  it was enhanced in \textit{s}-process elements via the third dredge-up of nucleosynthetic products created in the stellar interior and shell helium-burning (via thermal pulses), followed by subsequent mixing that can result in an over-abundance of C and N at the surface \citep{Karakas2014}. Moreover, if the extra-tidal star  is an actual evolved object, possibly an early-AGB or an AGB star, it would  have a mass in the range $\sim 2.5$ to $\sim3.7$ M$_{\odot}$ to explain the observed photospheric chemical composition.  As pointed out above, this would imply that a GC origin of this star is ruled out. 
	
	\item[$\bullet$] A third possibility could be that our star  is an \textit{s}-process-enriched red giant that is the result of star formation in an {\it already} \textit{s}-process-enriched medium, where AGB stars have played a more dominant role in chemical evolution \citep{Ventura2016}. Hence, a valid possibility is that the extra-tidal star inherited its present chemical composition while it was still a cluster member, likely from gas lost by a previous generation of $\sim$2.5--3.7 M$_{\odot}$ AGB stars.  This mass range is also fairly well-supported by the observed [Mg/Fe], and [Al/Fe] abundance ratios, as shown in Figure \ref{Figure4}.
	
	\item[$\bullet$] A fourth plausible interpretation is that N-rich stars towards the bulge region are among the oldest in the Galaxy, and their abundances are in fact the imprints of the very early chemical enrichment by spinstars, metal-poor, fast-rotating massive stars \citep{Chiappini2011, Barbuy2014}, which polluted the interstellar medium from which bulge GCs formed. An enhancement in N and modest enhancement in C, as well as some contribution to the neutron-capture elements (\textit{s}-process nucleosynthesis) might be expected. This scenario would predict that all of the stars in NGC~6723 (and indeed in all bulge GCs formed in situ) would be N-rich stars, whereas if the third possibility is correct, then there should exist first-generation stars in the cluster that do not exhibit enhanced nitrogen.
	
\end{itemize}

We conclude that it is very likely that we have identified a former cluster member (the extra-tidal star), with observed abundances consistent with production by AGB stars that played an important role in the chemical enrichment of NGC~6723, which is in-line with previous results for GCs at similar metallicity \citep{Ventura2016, Meszaros2020}. 

It is important to note that, at [Fe/H] $\sim -1.17$, it seems likely that the \textit{s}-process would dominate the production of such elements, invoking the AGB progenitor \citep{Kobayashi2020} rather that other possible sources such as binary neutron star (NS–NS) mergers \citep{Wanajo2014}. In addition, if we were to assume that the \textit{n}-capture came from an NS-NS merger, then we would need a second progenitor to account for the N and other light elements.

We also note some differences between the extra-tidal star and MW field stars. Figure \ref{Figure5} shows a comparison between this star (red symbol) and bulge field stars at similar metallicity from the APOGEE-2 survey. While this result reinforces that N-rich stars of different generations \citep{Martell2016} populate different regions of the canonical abundance planes, some similarities are noteworthy with Galactic field stars. The unique nature of our star is clear in the [C/Fe]--[N/Fe], [N/Fe]--[Al/Fe], [O/Fe]--[Al/Fe], [Mg/Fe]--[Al/Fe], [Mg/Fe]--[Si/Fe], [Mg/Fe]--[K/Fe], and [Ca/Fe]--[Ti/Fe] planes, which slightly exceed the Galactic levels, while some similarities remain for the $\alpha-$elements Mg, Si, Ca, and Ti, the odd-Z elements Al and K, and the iron-peak element Ti. In contrast to bulge field stars, the extra-tidal star clearly has higher [C, N, O/Fe] abundance ratios ($>+0.5$). These differences (and similarities) suggest that the extra-tidal star exhibits the typical chemical patterns similar to those of evolved stars in GC at [Fe/H]$\sim-1.0$ \citep{Meszaros2020}, as shown in panel (e) of Figure \ref{Figure3}. However, it is important to note that our extra-tidal star shows a lower metallicity compared to the cluster mean, but all its chemical patterns show good agreement with the other members of NGC~6723. This low-metallicity effect could be explained if some variability signature is detected in this star, as there is some evidence in the literature \citep{Munoz2018} that variability affects the measurement of the iron abundance in some way, causing a large offset with respect to the cluster mean. Another detailed photometric/spectroscopic analysis of this star is needed to investigate possible variability effects on the metallicity derivation. We note that, according to its radial velocity, proper motions, position in the CMD, and atmospheric parameters, this star is/was very likely a cluster member.

\section{Expected Radial Profile of halo field stars with a globular cluster-like abundance patterns toward NGC~6723}

We compute the predicted number ($N_{gc}$) of halo field stars with globular cluster-like abundance patterns observed in APOGEE-2 \citep[see, e.g.,][]{Fernandez-Trincado2016a, Fernandez-Trincado2017, Schiavon2017, Fernandez-Trincado2019c} towards the field of NGC~6723 using the smooth halo density relations presented in \citet{Horta2021}, and by adopting the same Monte Carlo implementation of the Von Neumann Rejection Technique \citep{Press2002} as in Eq. 1 in \citet[][]{Mateu2009} and Eq. 7 in \citep[][]{Fernandez-Trincado2015b}.

We find the expected number of observed APOGEE-2 halo field stars with a possible GC origin in the range $5.5$ kpc $<d_{\odot}< 9.5$ kpc, over a circular sky area of one-degree radius centred in NGC~6723, and with astrometric and kinematic properties as the cluster to be $N_{gc}< 0.05$ (from 1000 Monte Carlo realisations). This indicates that the region around NGC~6723 is underdense, not overdense, in Galactic halo field stars with globular cluster-like abundance patterns, leaving enough room to favour the scenario of a genuine extra-tidal star associated with NGC~6723.   

\section{Galactic orbit}
\label{section6}

We used the \texttt{GravPot16}\footnote{\url{https://gravpot.utinam.cnrs.fr}} algorithm for the calculation of the orbital path of both NGC~6723 and the extra-tidal star around the MW. The \texttt{GravPot16} model is composed of multiple potentials for the Galactic disc and a box/peanut bulge. For the Sun's position in the MW we assumed a distance to the Galactic Centre of $R_{\odot}=$8 kpc. In order to correct for the solar motion, we used the Sun's velocity respect to the local standard of rest ($U_{\odot}$, $V_{\odot}$, $W_{\odot}$) $= $($11.10, 12.24, 7.25$) km s$^{-1}$ \citep{Brunthaler2011} and the velocity of the local standard of rest (LSR) $\nu_{LSR}= 244.5$ km s$^{-1}$, based on $R_{\rm o}$ and assuming the composite rotation curve of \citet{Sofue2015}. 

For NGC~6723 and the extra-tidal star, we integrated over a 2 Gyr timespan, and calculated their orbital paths and orbital parameters using three different values of the angular velocity of the bar $\Omega_{\rm bar} = 33, 43, $ and 53 km s$^{-1}$ kpc$^{-1}$.  For NGC~6723, we adopted the observational parameters from \citet{Baumgardt2019}; for the extra-tidal star, we adopted the observational parameters given in Table \ref{table1}, except for the distance, for which we assumed two different values, e.g., $8.3\pm0.83$ kpc, the same heliocentric distance as NGC~6723 \citep{Baumgardt2019}, and $6.24\pm1.69$ kpc, the estimated distance from the \texttt{StarHorse} code \citep{Anders2019} (the algorithm combines the \textit{Gaia} astrometric information with multiband photometric information and a number of Galactic priors), which have smaller uncertainties at large ranges, as previously expected, and which have been proven to be an useful tool in confirming the dichotomy with APOGEE-2/DR16 and dissecting the bulge stellar populations \citep{Queiroz2018, Queiroz2020}.  

A simple Monte Carlo analysis was performed for the complete calculation, taking into account the errors in distance, RV, and proper motions, which were randomly propagated as 1$\sigma$ variations, assumed to follow  Gaussian distributions. For both objects we computed 1000 orbits. Figure \ref{Figure8} shows the probability densities of the resulting orbits on the X--Y, X--Z, and Y--Z planes in the non-inertial reference frame where the bar is at rest, where the dark area correspond to the most probable regions of the space, which are crossed more frequently by the simulated orbits. We found that both cluster and extra-tidal star are situated in the inner region of the bulge. For the full ensemble of orbits, we also calculated the deviations in the orbital parameters due to the different values of the angular velocity of the bar, ($\Delta{}r_{peri}$, $\Delta{}r_{apo}$, $\Delta{}Z_{max}$, $\Delta{}e$): NGC~6723 (0.02 kpc, 0.65 kpc, 0.37 kpc, 0.01), extra-tidal star at $d_{\odot}=8.3$ kpc (0.01 kpc, 0.52 kpc, 0.29 kpc, 0.01), and extra-tidal star at $d_{\odot}=6.24$ kpc (0.06 kpc, 0.07 kpc, 0.05 kpc, 0.03). With the different values of $\Omega_{\rm bar}$, we can see that the predicted orbits are not significantly affected by the change of this parameter, and consequently, it does alter our conclusions.  

\begin{figure*}
	\centering
	\includegraphics[width=185mm]{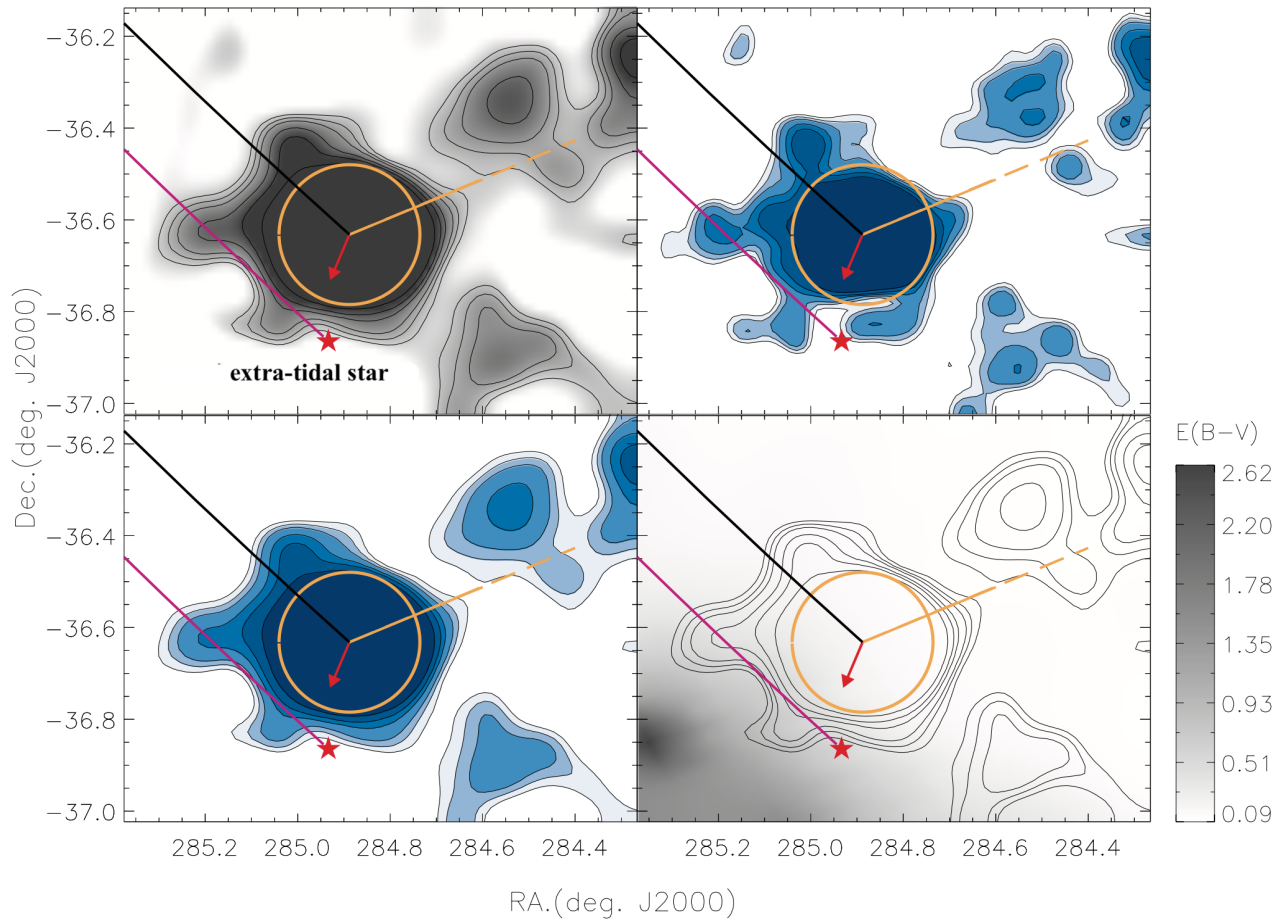}
	\caption{Modified Figure 10 from \citet{Chun2015}. The red `star' symbol shows the position of the newly identified extra-tidal star. The orange circle indicates the cluster tidal radius ($r_{t}= 9.14\pm0.49$ \arcmin, see Section \ref{section7}), and the red arrow defines the proper motion of NGC~6723 from \citet{Baumgardt2019}. The orange solid and dashed line shows the direction of the Galactic Centre and Galactic plane, respectively. The top-left panel is the star-count map around the cluster; the top-right panel is the surface-density map, smoothed by a Gaussian kernel 0.07$^{\circ}$; the lower-left panel is the same data, but smoothed with  a 0.11$^{\circ}$ kernel; the bottom-right panel shows the distribution map of E(B-V). The iso-density contour levels shown are 2.0$\sigma$, 2.5$\sigma$, 3.0$\sigma$, 4.0$\sigma$, 5.0$\sigma$, and 7.0$\sigma$. The orbital path of the cluster (black line) and the extra-tidal star (purple line) is over-plotted.}  
	\label{Figure7}
\end{figure*}	

The median orbital parameters, assuming $\Omega_{\rm bar}=43$ km s$^{-1}$ kpc for NGC~6723 and the extra-tidal star, are as follows: ($r_{peri}$, $r_{apo}$, $Z_{max}$, $e$): NGC~6723 ($0.05\pm0.08$ kpc, $3.28\pm0.12$ kpc, $3.87\pm0.12$ kpc, $0.97\pm0.04$), extra-tidal star at $d_{\odot}=8.3$ kpc ($0.06\pm0.14$ kpc, $3.36\pm0.10$ kpc, $3.98\pm0.12$ kpc, $0.97\pm0.07$), and extra-tidal star at $d_{\odot}=6.24$ kpc ($1.56\pm1.07$ kpc, $4.28\pm0.66$ kpc, $2.46\pm0.82$ kpc, $0.46\pm0.24$). We confirm that both NGC~6723 and the newly found extra-tidal star possess trajectories indicating that they are confined to the bulge region ($\lesssim4$ kpc) and have similar orbital properties, suggesting that the extra-tidal star shares an identical dynamical history as NGC~6723, as expected of stripped cluster stars. Figure \ref{Figure1} (b) and Figure \ref{Figure7} show that the extra-tidal star is in-line with the cluster's orbit. Figure \ref{Figure7} also suggests that the orbital path of NGC~6723 is tracing out the trailing arm, and our extra-tidal star candidate is located along the leading arm. 

\begin{figure}
	\centering
	\includegraphics[width=90mm]{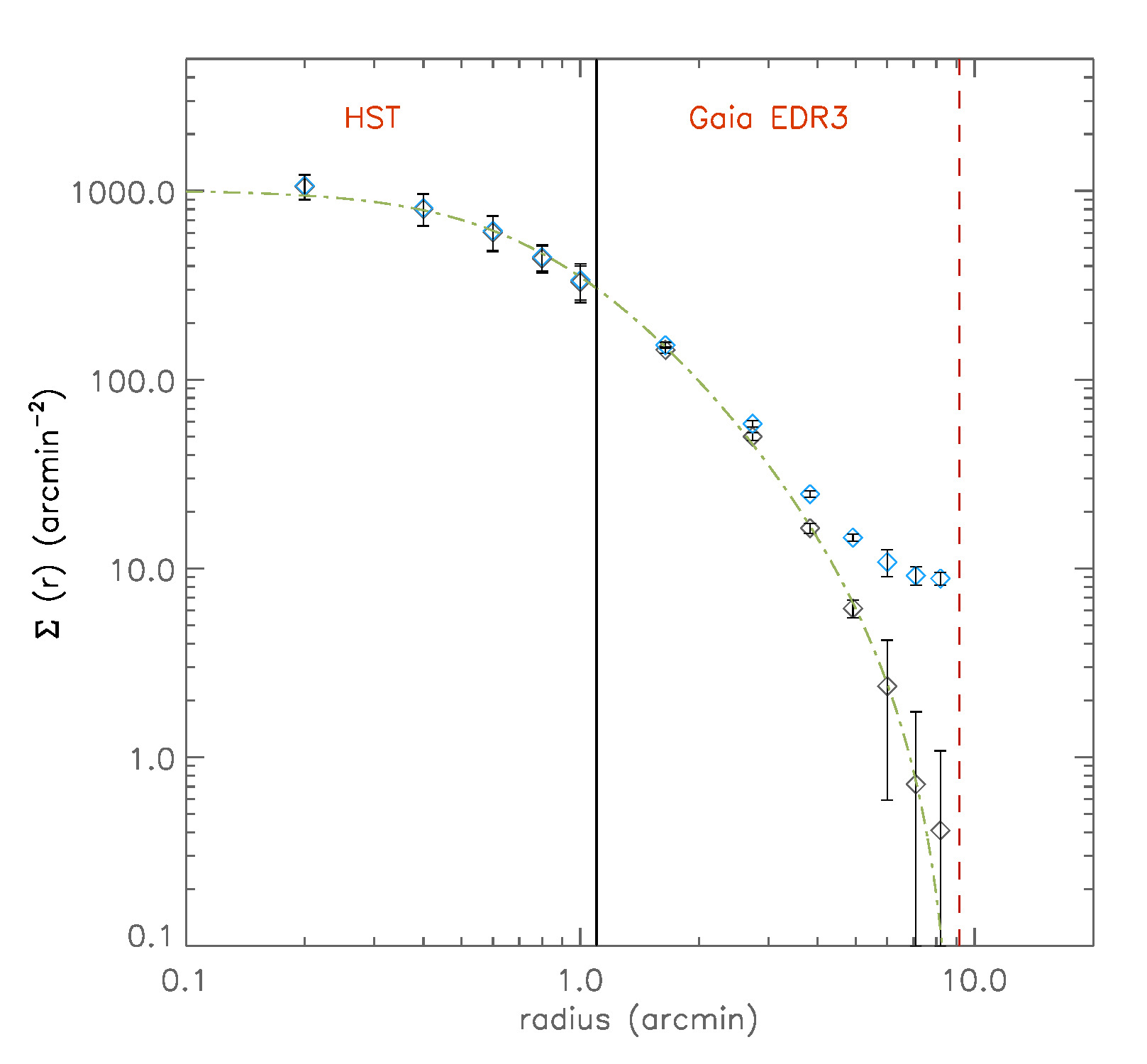}
	\caption{Stellar-density profile of NGC~6723 determined from the HST$+$\textit{Gaia} EDR3 data set. Small orange symbols show densities before background subtraction. The-best fit \citet{King1966} profile is shown in lime and cyan, respectively. The red dashed line show the determined tidal radius ($r_{t}=9.14\pm0.49$ \arcmin) in this work, while the black line in 1.1 \arcmin separate the HST  from \textit{Gaia} EDR3 data set. }  
	\label{Figure9}
\end{figure}	

\section{Comments on the tidal radius\\}
\label{section7}

The tidal radius for NGC~6397 has been measured by several studies, resulting in different values. For instance, a recent work by \citet{deBoer2019} using \textit{Gaia} DR2 data determined a tidal radius that peaks at $r_{t}= 67.66 \pm9.50$ pc ($\sim26.7'$) from a generalised lowered isothermal \texttt{LIMEPY} model. This is considerably larger than previously thought. The \citet{Baumgardt2019} value, using the same data set and a different analysis method, adopted $r_{t}= 36.95$ pc ($\sim$14.6$'$), which is substantially larger than the value of $r_{t}=10.51'$ listed by \citet[][2010 edition]{Harris1996, Harris2010}. 

Moreover, simulations from \citet{Moreno2014} calculated tidal radii between 29 -- 31.4 pc using a 6-D dataset in an axisymmetric and non-axisymmetric Galactic potential, including a rotating Galactic bar and a 3-D model for the spiral arms. Their results indicate good agreement with the tabulated value from Harris's compilation and the cluster Jacobi radii determined in \citet{Piatti2019}. Furthermore, the observed cluster stellar-density maps from near-infrared observations \citep[see, e.g.,][]{Chun2015} suggest the existence of over-density features of some number of stars from $\sim 6.3'$ $\lesssim r \lesssim 15.77'$, exhibiting azimuthally irregular patterns. Our newly identified extra-tidal star lies in this region as shown in Figure \ref{Figure7}, and placed well outside the Jacobi radius \citep[$\gtrsim11\arcmin$;][]{Piatti2019}. This called our attention to the recent derived slightly larger $r_{t}$ values from \citet[][]{deBoer2019}, based on \textit{Gaia} DR2. This value, should be viewed with caution, as it has its own limitations toward the bulge regions, and could lead to fitting non-physically motivated core \cite{King1962} and tidal radii. 

With the new improved data from \textit{Gaia} EDR3 \citep{Brown2020}, and the available \textit{Hubble Space Telescope} (HST) imaging data within 1.1 \arcmin from the cluster center \citep{Sarajedini2007}, we can determine updated structural parameters with better precision. To construct number density profiles, we make use of star counts by adopting the same technique as in \citet{Cohen2020} to combine both surveys. Briefly, the stellar density is found in annuli of varying radii around the centre of the cluster, and these density values are fitted by a King profile \citep{King1966}. The HST$+$\textit{Gaia} EDR3 data is fairly well-fit by the \citep{King1966} profile. The stellar-density profile is shown in Figure \ref{Figure9}. For NGC~6723 we have determined a radius $r_{t}=9.14\pm0.49$ \arcmin, which is in good agreement with the values reported in \citet[][2010 edition]{Harris1996}, and by dynamical studies of NGC~6723 \citep[see, e.g.,][]{Moreno2014, Piatti2019}.

Our finding is in line with recent studies in the vicinity of NGC~6723, for example, near-infrared J-, H-, and K-band observations from the WFCAM camera on the 3.8 m UKIRT \cite{Chun2015}, suggesting the existence of extra-tidal features $\gtrsim11'$ (Fig. \ref{Figure7}), and is in good agreement with a rather high destruction ratio of $\nu_{tot}/\nu_{evap}\sim0.81 - 2.0$ \citet{Gnedin1997, Moreno2014} (where $\nu_{tot}$ is the total destruction rate of this cluster, and $\nu_{evap}$ is the evaporation rate per Hubble time). The newly identified extra-tidal star lie near the weak extended substructure beyond the tidal radius in the southern region of NGC~6723 \citep[see, e.g.,][]{Chun2015} as highlighted in Figure \ref{Figure7}, suggesting that this star has left the GC's potential. Thus, our findings provide strong support for the idea of linking N-rich field stars to GCs in the bulge region of the MW \citep{Bekki2019}.

\section{Concluding remarks}
\label{section8}

Our finding is in-line with recent studies in the vicinity of NGC~6723, e.g., near-infrared J-, H-, and K-band observations from the WFCAM camera on the 3.8 m UKIRT \citep{Chun2015}, suggesting the existence of extra-tidal features around this cluster (see Figure~\ref{Figure7}). Thus, our findings provide strong support for the idea of linking some of the observed N-rich field stars to GCs in the bulge region of the MW \citep[e.g.,][]{Schiavon2017, Fernandez-Trincado2019c}.

We also conclude that it is very likely that we have identified a former cluster member, with observed abundances consistent with production by AGB stars that played an important role in the chemical enrichment of NGC~6723, which is in-line with previous results for GCs at similar metallicity \citep{Meszaros2020}.

There are also several competing scenarios that might explain the unusual chemical abundances of the newly identified extra-tidal star, in the context of an escaped member of NGC~6723 that was tidally disrupted and captured by the MW's bulge.  In summary:

\begin{itemize}
	
	\item[$\bullet$] \textit{Extrinsic mechanism}: The over-abundance of the \textit{s}-process elements in this star could come from the accretion of \textit{s}-process-rich matter from a former thermally pulsing AGB companion during its heavy mass-loss phase on the AGB, which has since evolved into a faint white dwarf star. 
	
	\item[$\bullet$] \textit{Intrinsic mechanism}: This star could be an intrinsic AGB star, enhanced in \textit{s}-process elements during the third dredge-up of nucleosynthetic products created in the stellar interior and shell helium-burning (via thermal pulses), followed by subsequent mixing that can result in an over-abundance of carbon and nitrogen at the surface.
	
	\item[$\bullet$] This star could be an \textit{s}-process-enriched red giant resulting from star formation in an already \textit{s}-process-enriched medium, where low-mass AGB stars have played a dominant role in chemical evolution.
	
\end{itemize}

Finally, the newly identified extra-tidal star displays a [N/Fe] ratio similar to other metal-poor N-rich field stars, but which are not carbon enriched (one of the typical signatures of second-generation GC stars)--\citep[see e.g.,][]{Fernandez-Trincado2016a, Fernandez-Trincado2017, Schiavon2017, Fernandez-Trincado2019c}, suggesting that the extra-tidal star is not a genuine second-generation star, but it is part of a sub-family of the N-rich stars with modest carbon enrichment as that observed in some Galactic GC stars at similar metallicity, likely associated with the intermediate-mass ($\lesssim$3 M$_{\odot}$) population of early-AGB stars. 

   \begin{sidewaystable*}
	\begin{small}
		\setlength{\tabcolsep}{0.80mm}  
			\caption{\textit{Top table:} Final abundances for the stars analyzed in this work , adopting stellar atmospheric parameters from photometry (T$_{\rm eff}$) and log $g$ from 12.5 Gyr \texttt{PARSEC} isochrones, as determined using local thermodynamic equilibrium (LTE) model atmospheres. \textit{Middle table}: The final abundances as determined by the \texttt{ASPCAP} pipeline. \textit{Bottom table}: List of the main physical parameters of our stars. }
			\centering
			\begin{tabular}{|c|cccccccccccccccccc|}
				\hline
				APOGEE$\_$ID           &     T$_{\rm eff}$  & log $g$   &     [M/H]    &     $\xi_{t}$  &      [C/Fe]   &      [N/Fe]  &      [O/Fe]  &      [Mg/Fe]  &      [Al/Fe]  &      [Si/Fe]  &      [K/Fe]   &      [Ca/Fe]  &      [Ti/Fe]  &      [Fe/H]   &      [Ni/Fe]  &      [Ce/Fe]  &      [Nd/Fe]  &      [Yb/Fe]  \\
				{\bf This work}  &  (K) &  & & km s$^{-1}$&       &     &   &      &     &     &   &     &       &      &        &      &      &   \\				
				\hline
				2M18590940$-$3640192  &     4397       &     $1.28$   &     $-1.11$  &     1.766      &      $-0.45$  &      $+1.02$  &      $+0.46$  &      $+0.37$   &      $+0.16$   &      $+0.42$   &      $-0.01$  &        $+0.09$   &      $+0.29$   &      $-0.93$  &      $-0.06$  &      $+0.04$   &      ...      &      ...      \\
				2M18592786$-$3636509  &     4425       &     $1.41$   &     $-0.96$  &     1.931      &      $-0.58$  &      $+1.53$  &      $+0.55$  &      $+0.32$   &      $+0.40$   &      $+0.29$   &      $+0.19$   &      $+0.30$   &      $+0.47$   &      $-0.91$  &      $-0.07$  &      $+0.29$   &      ...      &      ...      \\
				2M18593255$-$3641374  &     4349       &     $1.27$   &     $-1.02$  &     1.817      &      $-0.17$  &      $+0.56$  &      $+0.60$  &      $+0.30$   &      $+0.24$   &      $+0.35$   &      $+0.21$   &      $+0.36$   &      $+0.19$   &      $-1.02$  &      $+0.01$   &    $+0.15$   &      $+0.22$   &      ...      \\
				2M18593691$-$3641280  &     4498       &     $1.62$   &     $-0.87$  &     1.549      &      $-0.20$  &      $+0.86$  &      $+0.14$  &      $+0.33$   &      $+0.52$   &      $+0.37$   &      $+0.14$   &      $+0.21$   &      $+0.08$   &      $-1.00$  &      $+0.05$   &    $+0.11$   &      ...      &      ...      \\
				2M18594472$-$3639043  &     4364       &     $1.30$   &     $-1.05$  &     1.761      &      $-0.53$  &      $+1.06$  &      $+0.16$  &      $+0.31$   &      $+0.19$   &      $+0.41$   &      $+0.02$   &      $+0.14$   &      $+0.19$   &      $-0.99$  &      $-0.09$  &      $+0.05$   &      ...      &      ...      \\
				2M18594565$-$3637016  &     4056       &     $0.74$   &     $-1.01$  &     2.046      &      $-0.21$  &      $+0.45$  &      $+0.47$  &      $+0.31$   &      $+0.24$   &      $+0.35$   &      $+0.21$   &      $+0.32$   &      $+0.19$   &      $-1.02$  &      $+0.03$   &    $+0.13$   &      $+0.07$   &      $+0.59$   \\
				2M18594898$-$3635401  &     4054       &     $0.74$   &     $-0.97$  &     2.812      &      $-0.29$  &      $+1.18$  &      $+0.39$  &      $+0.10$   &      $+0.21$   &      $+0.31$   &      $+0.31$   &      $+0.35$   &      $+0.45$   &      $-1.12$  &      $+0.03$   &    $+0.22$   &      ...      &      $+0.76$   \\
				\hline
				\hline
				2M18594405$-$3651518  &     4130       &     $0.80$   &     $-1.09$  &     2.733      &      $+0.55$   &      $+0.94$  &      $+0.54$  &      $+0.18$   &      $+0.17$   &      $+0.38$   &      $+0.33$   &      $+0.24$   &      $+0.37$   &      $-1.17$  &      $+0.11$   &      $+1.19$   &      $+1.54$   &      $+1.29$   \\
				\hline
				\hline
				APOGEE$\_$ID           &     T$_{\rm eff}$  &  log $g$   &     [M/H]    &     $\xi_{t}$  &      [C/Fe]   &      [N/Fe]  &      [O/Fe]  &      [Mg/Fe]  &      [Al/Fe]  &      [Si/Fe]  &      [K/Fe]   &      [Ca/Fe]  &      [Ti/Fe]  &      [Fe/H]   &      [Ni/Fe]  &      [Ce/Fe]  &      [Nd/Fe]  &      [Yb/Fe]  \\
				{\bf \texttt{ASPCAP} } &  (K) & && km s$^{-1}$ &    &     &   &    &     &       &  &    &     &       &       &   & &    \\							
				\hline
				2M18590940$-$3640192  &     4377       &     $1.32$   &     $-1.11$  &     2.503      &      $-0.22$  &      $+0.71$  &      $+0.28$  &      $+0.28$   &      $-0.03$  &      $+0.26$   &      $+0.10$   &      $+0.16$   &       $+0.06$    &      $-1.11$  &      $-0.03$  &      $-0.07$  &      ...      &      ...      \\
				2M18592786$-$3636509  &     4391       &     $1.65$   &     $-0.96$  &     2.126      &      $-0.24$  &      $+0.99$  &      $+0.22$  &      $+0.24$   &      $+0.19$   &      $+0.23$   &      $+0.24$   &      $+0.23$   &      $+0.31$   &      $-0.98$  &      $+0.01$   &      $+0.24$   &      ...      &      ...      \\
				2M18593255$-$3641374  &     4327       &     $1.53$   &     $-1.03$  &     1.849      &      $-0.06$  &      $+0.09$  &      $+0.28$  &      $+0.24$   &      $+0.02$   &      $+0.31$   &      $+0.25$   &      $+0.22$   &      ...              &      $-1.02$  &      $+0.04$   &      $+0.00$   &      ...      &      ...      \\
				2M18593691$-$3641280  &     4847       &     $2.29$   &     $-0.88$  &     2.017      &      $-0.20$  &      $+1.01$  &      $+0.14$  &      $+0.18$   &      $+0.48$   &      $+0.22$   &      $+0.27$   &      $+0.26$   &      $+0.42$   &      $-0.90$  &      $+0.04$   &      $+0.49$   &      ...      &      ...      \\
				2M18594472$-$3639043  &     4597       &     $1.58$   &     $-1.05$  &     2.545      &      $-0.26$  &      $+0.88$  &      $+0.22$  &      $+0.22$   &      $+0.05$   &      $+0.22$   &      $+0.14$   &      $+0.13$   &      $+0.23$   &      $-1.06$  &      $+0.01$   &      $+0.30$   &      ...      &      ...      \\
				2M18594565$-$3637016  &     4121       &     $1.14$   &     $-1.01$  &     2.025      &      $-0.13$  &      $+0.17$  &      $+0.28$  &      $+0.33$   &      $+0.05$   &      $+0.28$   &      $+0.24$   &      $+0.24$   &      ...      &      $-1.00$  &      $+0.02$   &      ...      &      ...      &      ...      \\
				2M18594898$-$3635401  &     4204       &     $1.31$   &     $-0.98$  &     2.123      &      $-0.24$  &      $+0.94$  &      $+0.22$  &      $+0.25$   &      $+0.21$   &      $+0.21$   &      $+0.29$   &      $+0.28$   &      ...      &      $-0.99$  &      $+0.01$   &      $+0.13$   &      ...      &      ...      \\
				\hline
				\hline
				2M18594405$-$3651518  &     4058       &     $-0.19$  &     $-1.09$  &     1.995      &      $+0.04$   &      $+0.82$  &      $+0.04$  &      $+0.33$   &      $+0.11$   &      $+0.19$   &      $+0.30$   &      $+0.18$   &      ...      &      $-1.15$  &      $-0.13$  &      ...      &      ...      &      ...      \\
				\hline
				\hline
				APOGEE\_ID &   S/N &   $RV$  &   $\sigma RV$ &  E(B-V) &  V$_{\rm o}$ & (B-V)$_{\rm o}$ & H$_{\rm o}$ & (J-K$_{s}$)$_{\rm o}$ &  $\mu_{\alpha} \cos(\delta)$ & $\mu_{\delta}$ &   \texttt{RUWE} & \# Visits && & & & & \\		
				&  & km s$^{-1}$ &   km s$^{-1}$      &    &        &    &  &  &   & (mas yr$^{-1}$) & (mas yr$^{-1}$)   &  & &&& &  & \\				    
				&   &     &     &   \texttt{CTIO}  & \texttt{CTIO}  & \texttt{2MASS}  & \texttt{2MASS}  & &    &  & & & &&&&&\\				    
				\hline 
				2M18590940$-$3640192 &    128 &   $-$96.5   &   0.01  &     0.063  & 13.34  & 1.19  & 10.51  &  0.76    &  0.92$\pm$0.02 & $-$2.47$\pm$0.01  &   1.18    & 2 &&&&& & \\
				2M18592786$-$3636509 &    187 &   $-$91.3   &   0.00  &     0.063  & 13.45  & 1.25  & 10.45  &  0.75    &  0.84$\pm$0.02 & $-$2.51$\pm$0.02  &   1.16    & 2 &&&&& & \\
				2M18593255$-$3641374 &    123 &   $-$89.5   &   0.04  &     0.063  & 13.69  & 1.19  & 10.83  &  0.78    &  0.89$\pm$0.02 & $-$2.28$\pm$0.02  &   1.18    & 2 &&&&& & \\
				2M18593691$-$3641280 &    98   &   $-$91.4   &   0.11  &     0.063  & 14.52  & 1.02  & 11.86  &  0.72    &  1.04$\pm$0.03 & $-$2.37$\pm$0.02  &   1.10    & 2 &&&&& & \\
				2M18594472$-$3639043 &    165 &   $-$100.1 &   0.01  &     0.063  & 13.54  & 1.13  & 10.81  &  0.77    &  1.25$\pm$0.02 & $-$2.45$\pm$0.02  &   1.20    & 2 & &&&& &\\
				2M18594565$-$3637016 &    258 &   $-$91.9   &   0.29  &     0.063  & 13.22  & 1.37  & 10.01  &  0.89    &  0.84$\pm$0.02 & $-$2.53$\pm$0.02  &   1.13    & 2 &&&&& &\\
				2M18594898$-$3635401 &    281 &   $-$93.7   &   ...      &     0.063  & 13.08  & 1.43  &  9.77   &  0.91    &  0.93$\pm$0.02 & $-$2.55$\pm$0.02  &   1.21    & 2 &&&&& &\\
				\hline
				\hline
				2M18594405$-$3651518 &    270 &   $-$94.5   &   0.06  &     0.520  & 12.72  & 1.38  &  9.78   &  0.86    &  1.06$\pm$0.02 & $-$2.32$\pm$0.02  &   1.19    & 2 &&&&& &\\
				\hline
			\end{tabular}  \label{table1}\\
	\tablefoot{{\bf Note:} The synthetic spectra were based on 1D Local Thermodynamic Equilibrium (LTE) model atmospheres calculated using \texttt{MARCS} models \citep{Gustafsson2008} and the Solar abundances from \citet{Asplund2005}, except for Ce II, Nd II, and Yb II, for which we have adopted the Solar abundances from \citet{Grevesse2015}.}
	\end{small}
\end{sidewaystable*}

\begin{table}
	\begin{center}
		\setlength{\tabcolsep}{0.4mm}  
		\caption{Abundance-ratio uncertainties for the model parameters with $\Delta$T$_{\rm eff} = 100$ K, $\Delta \log g$ $=$ 0.3 dex, $\Delta\xi_{t} = 0.05$ km s$^{-1}$, the uncertainties on the mean due to line-to-line scatter ( $\sigma_{\rm [X/H], mean}$), and corresponding total error.}
		\begin{tabular}{|l|ccccc|}
			\hline
			{\bf 2M18594405}     &  $\sigma_{\rm [X/H],\log g}$    &   $\sigma_{\rm [X/H],T_{\rm eff}}$   &   $\sigma_{\rm [X/H],\xi_{t}}$   &   $\sigma_{\rm [X/H], mean}$ &   $\sigma_{\rm [X/H], total}$    \\
			{\bf $-$3651518}     &     &    &   &   &      \\
			& [dex]  &  [dex]  & [dex] & [dex] & [dex] \\
			\hline
			\hline
			$^{12}$C$^{16}$O    $\rightarrow$ $^{12}$C   &  0.065 &  0.018 &  0.004 & 0.110 &  0.129 \\
			$^{12}$C$^{14}$N    $\rightarrow$ $^{14}$N  &  0.075 &  0.094 &  0.023 & 0.071 &  0.142 \\
			$^{16}$OH  $\rightarrow$   $^{16}$O  &  0.018 &  0.097 &  0.014 & 0.068 &  0.121 \\
			Mg I  &  0.005 &  0.089 &  0.022 & 0.052 &  0.106 \\
			Al I  &  0.011 &  0.080 &  0.006 & 0.117 &  0.142 \\
			Si I  &  0.045 &  0.023 &  0.006 & 0.094 &  0.107 \\
			K I  &  0.025 &  0.044 &  0.009 & 0.137 &  0.146 \\
			Ca I  &  0.023 &  0.023 &  0.007 & 0.085 &  0.091 \\
			Ti I   &  0.007 &  0.070 &  0.033 & 0.246 &  0.258 \\
			Fe I  &  0.016 &  0.042 &  0.005 & 0.154 &  0.161 \\
			Ni I   &  0.027 &  0.047 &  0.010 & 0.155 &  0.165 \\
			Ce II    &  0.123 &  0.067 &  0.025 & 0.125 &  0.189 \\
			Nd II  &  0.075 &  0.071 &  0.023 & 0.130 &  0.168 \\
			Yb II   & 0.092  &  0.024 &  0.004 & 0.020 &  0.097  \\
			\hline
			\hline
			{\bf 2M18594898}  &  $\sigma_{\rm [X/H],\log g}$    &   $\sigma_{\rm [X/H],T_{\rm eff}}$   &   $\sigma_{\rm [X/H],\xi_{t}}$   &   $\sigma_{\rm [X/H], mean}$    &   $\sigma_{\rm [X/H], total}$    \\
			{\bf $-$3635401}  &     &    &   &   &      \\
			& [dex]  &  [dex]  & [dex] & [dex] & [dex] \\
			\hline
			\hline
			$^{12}$C$^{16}$O    $\rightarrow$ $^{12}$C        &    0.035   &    0.064   &    0.009   &       0.052   &     0.090 \\ 
			$^{12}$C$^{14}$N    $\rightarrow$ $^{14}$N          &    0.042   &    0.108   &    0.004   &       0.069   &     0.135 \\
			$^{16}$OH  $\rightarrow$   $^{16}$O        &    0.049   &    0.139   &    0.003   &       0.035   &     0.152  \\
			Mg  I  &    0.025   &    0.073   &    0.016   &       0.017   &     0.081 \\
			Al I    &    0.014   &    0.111   &    0.009   &       0.116   &     0.161  \\
			Si I     &    0.033   &    0.007   &    0.003   &       0.079   &     0.086  \\
			K I      &    0.038   &    0.009   &    0.004   &       0.014   &     0.042   \\
			Ca I    &    0.037   &    0.033   &    0.003   &       0.026   &     0.056  \\
			Ti I     &    0.067   &    0.070   &    0.043   &       0.199   &     0.225  \\
			Fe I     &    0.036   &    0.072   &    0.008   &       0.116   &     0.141 \\
			Ni I     &    0.011   &    0.021   &    0.003   &       0.070   &     0.074  \\
			Ce II    &    0.059   &    0.013   &    0.004   &       0.069   &     0.092  \\ 
			Nd II   &    0.049   &    0.043   &    0.005   &       0.043   &     0.078  \\
			Yb II    &    0.150   &    0.027    &   0.002   &       0.047   &     0.159  \\
			\hline
		\end{tabular}  \label{table2}\\
	\end{center}
	\raggedright{{\bf Note:} The total error is defined as: $\sigma_{\rm [X_{\rm i}/Fe], total}  = \sqrt{\sigma^2_{\rm [X_{\rm i}/Fe], \log g} + \sigma^2_{\rm [X_{\rm i}/Fe],  T_{\rm eff}}  +  \sigma^2_{\rm [X_{\rm i}/Fe], \xi_t}  + \sigma^2_{\rm [X_{\rm i}/Fe], mean} }$.}
\end{table}

\begin{acknowledgements}  
We thank the anonymous referee for helpful comments that greatly improved the paper.	
We thank Dr. Lee Y.-W. 
J.G.F-T is supported by FONDECYT No. 3180210. 
T.C.B. acknowledge partial support for this work from grant PHY 14-30152; Physics Frontier Center / JINA Center for the Evolution of the Elements (JINA-CEE), awarded by the US National Science Foundation.
D.M. is supported by the BASAL Center for Astrophysics and Associated Technologies (CATA) through grant AFB 170002, and by project FONDECYT Regular No. 1170121. 
D.G. gratefully acknowledges support from the Chilean Centro de Excelencia en Astrof\'isica
y Tecnolog\'ias Afines (CATA) BASAL grant AFB-170002. D.G. also acknowledges financial support from the Direcci\'on de Investigaci\'on y Desarrollo de la Universidad de La Serena through the Programa de Incentivo a la Investigaci\'on de Acad\'emicos (PIA-DIDULS).
S.V. gratefully acknowledges the support provided by Fondecyt reg. n. 1170518. 
S.O.S. acknowledges the FAPESP PhD fellowship 2018/22044-3.
B.B. acknowledge partial financial support from FAPESP, CNPq, and CAPES - Finance Code 001. 
A.P-V. and S.O.S acknowledge the DGAPA-PAPIIT grant IG100319.
B.T. gratefully acknowledges support from the National Natural Science Foundation of China under grant No. U1931102 and support from the hundred-talent project of Sun Yat-sen University. 
J.A.-G. acknowledges support from Fondecyt Regular 1201490 and from ANID, Millennium Science Initiative ICN12\_009, awarded to the Millennium Institute of Astrophysics (MAS). We acknowledge UA VRIIP for the fund "Asistentes de Investigaci\'on" financed by projects ANT1855 and ANT1856 from Ministerio de Educaci\'on, Chile. 
T.P. acknowledge support from the Argentinian institutions CONICET and SECYT (Universidad Nacional de C\'ordoba). 
A.R-L acknowledges financial support provided in Chile by Agencia Nacional de Investigaci\'on y Desarrollo (ANID) through the FONDECYT project 1170476.
R.R.M. acknowledges partial support from project BASAL AFB-$170002$ as well as FONDECYT project N$^{\circ}1170364$. 
L.C-V thanks the Fondo Nacional de Financiamiento para la Ciencia, La Tecnolog\'ia y la innovaci\'on "FRANCISCO JOS\'E DE CALDAS", MINCIENCIAS, and the VIIS for the economic support of this research.

\newline
Funding for the Sloan Digital Sky Survey IV has been provided by the Alfred P. Sloan Foundation, the U.S. Department of Energy Office of Science, and the Participating Institutions. SDSS-IV acknowledges support and resources from the Center for High-Performance Computing at the University of Utah. The SDSS website is www.sdss.org. SDSS-IV is managed by the Astrophysical Research Consortium for the Participating Institutions of the SDSS Collaboration including the Brazilian Participation Group, the Carnegie Institution for Science, Carnegie Mellon University, the Chilean Participation Group, the French Participation Group, Harvard-Smithsonian Center for Astrophysics, Instituto de Astrof\`{i}sica de Canarias, The Johns Hopkins University, Kavli Institute for the Physics and Mathematics of the Universe (IPMU) / University of Tokyo, Lawrence Berkeley National Laboratory, Leibniz Institut f\"{u}r Astrophysik Potsdam (AIP), Max-Planck-Institut f\"{u}r Astronomie (MPIA Heidelberg), Max-Planck-Institut f\"{u}r Astrophysik (MPA Garching), Max-Planck-Institut f\"{u}r Extraterrestrische Physik (MPE), National Astronomical Observatory of China, New Mexico State University, New York University, the University of Notre Dame, Observat\'{o}rio Nacional / MCTI, The Ohio State University, Pennsylvania State University, Shanghai Astronomical Observatory, United Kingdom Participation Group, Universidad Nacional Aut\'{o}noma de M\'{e}xico, University of Arizona, University of Colorado Boulder, University of Oxford, University of Portsmouth, University of Utah, University of Virginia, University of Washington, University of Wisconsin, Vanderbilt University, and Yale University.

\newline
This work has made use of data from the European Space Agency (ESA) mission \textit{Gaia} (\url{http://www.cosmos.esa.int/gaia}), processed by the \textit{Gaia} Data Processing and Analysis Consortium (DPAC, \url{http://www.cosmos.esa.int/web/gaia/dpac/consortium}). Funding for the DPAC has been provided by national institutions, in particular the institutions participating in the \textit{Gaia} Multilateral Agreement.	
\end{acknowledgements}
	
\newpage	
	


\end{document}